\documentclass[preprint,aps,showpacs]{revtex4}

\usepackage{amsmath}
\usepackage{amssymb}
\usepackage{bbold}
\usepackage{mathrsfs}
\usepackage{graphicx, psfrag}
\usepackage[centerlast]{caption}
\usepackage{float}
\usepackage{subfig}
\usepackage{tikz}
\usepackage{pgfplots}
\usepackage[colorlinks=true, citecolor=blue, urlcolor = blue, linkcolor= red, 
bookmarks=true]{hyperref}
\usepackage{epstopdf}
\usepackage{multirow}
\usepackage{ulem}

\begin{document}
\def \beq{\begin{equation}}
\def \eeq{\end{equation}}
\def \bse{\begin{subequations}}
\def \ese{\end{subequations}}
\def \bea{\begin{eqnarray}}
\def \eea{\end{eqnarray}}
\def \bem{\begin{displaymath}}
\def \eem{\end{displaymath}}
\def \bem{\begin{pmatrix}}
\def \eem{\end{pmatrix}}
\def \bb{\bibitem}
\def \bs{\boldsymbol}
\def \nn{\nonumber}
\def \mf{\tilde{J}_1}
\def \mj{\tilde{J}_0}
\def \mh{\mathcal{H}}
\def \ma{\mathcal{A}}
\def \md{\mathcal{D}}
\def \mg{\mathcal{G}}
\def \ha{\hat{a}}
\def \hs{\hat{\xi}}
\def \hh{\hat{H}}
\def \mH{\hat{\mathcal{H}}}
\def \hp{\hat{\Psi}}
\def \hb{\hat{b}}
\def \hB{\hat{\mathcal{T}}}
\def \hN{\hat{\mathcal{N}}}

\newcommand{\cc}{\color{red}}
\newcommand{\cb}{\color{blue}}
\newcommand{\upa}{\uparrow}
\newcommand{\dna}{\downarrow}
\newcommand{\pdag}{\phantom\dagger}
\newcommand{\bra}[1]{\langle#1|}
\newcommand{\ket}[1]{|#1\rangle}
\newcommand{\braket}[2]{\langle#1|#2\rangle}
\newcommand{\ketbra}[2]{| #1 \rangle \langle #2 |}
\newcommand{\expect}[1]{\langle#1\rangle}
\newcommand{\inx}[1]{\int d\bs x \Big [ #1 \Big ]}
\newcommand{\si}[1]{\hp^{\pdag}_{#1} (\bs x)}
\newcommand{\dsi}[1]{\hp^\dag_{#1} (\bs x)}
\newcommand{\evolve}[1]{\frac{\partial #1}{\partial t}}


\title{\textbf{Ultracold Spin-Orbit Coupled Bose-Einstein Condensate  in a Cavity: Route to Magnetic Phases Through Cavity Transmission.}}
\author{ Bikash Padhi and Sankalpa Ghosh }
\affiliation{Department of Physics, IIT Delhi, New Delhi-110016, India}
\email{sankalpa@physics.iitd.ac.in}
\begin{abstract}
We study the spin orbit coupled ultra cold Bose-Einstein condensate placed in a single mode Fabry-P\'erot cavity. The cavity introduces a quantum optical lattice potential which dynamically couples with the atomic degrees of freedom and realizes a generalized extended Bose Hubbard model whose zero temperature phase diagram can be controlled by tuning the cavity parameters. In the non-interacting limit, where the atom-atom interaction is set to zero, the resulting atomic dispersion shows interesting features such as bosonic analogue of Dirac points,  cavity controlled Hofstadter spectrum which bears the hallmark of pseudo-spin-1/2 bosons in presence of Abelian and non-Abelian gauge field ( the later due to spin-orbit coupling) in a cavity induced optical lattice potential. In the presence of atom-atom interaction, using a mapping to a generalized Bose Hubbard model of spin-orbit coupled bosons in classical optical lattice, we show that the system realizes a host of quantum magnetic phases whose magnetic order can be be detected from the cavity transmission. This provides an alternative approach for detecting quantum magnetism in ultra cold atoms. We discuss the effect of cavity induced optical bistability on this phases and their experimental consequences. 

\end{abstract}
\maketitle
\section{Introduction}

Quantum Simulation of exotic condensed matter phases \cite{QMB, QS, Lewenstein-book} with ultra cold atoms witnessed tremendous progress in recent times. A significant step in the direction of realization of such exotic phases 
was taken through the experimental realization of synthetic spin-orbit coupling for bosonic ultra cold systems \cite{ Spielman1, Dali} and subsequently for fermionic ultracold atom \cite{FSO1, FSO2}. The development opened the possibility of simulating analogues of topologically non trivial condensed matter phases \cite{Hassan} as well as quantum magnetic phases \cite{Auerbach} in the domain of ultra cold atoms. All these development led to a flurry of theoretical as well as experimental activity in this direction \cite{reviewSOC}.  

In this work we consider such spin-orbit coupled (SOC) ultra cold Bose Einstein condensate (BEC) inside a Fabry-P\'erot cavity and study the consequences of atom-photon interaction on the phase diagram of SOC 
bosons. The motivation for studying the SOC ultra cold atoms in this unique environment have come from the recent progress in studying ultra cold atomic systems inside a high finesse single mode optical cavity 
\cite{RitschRMP, CavityExp1, Esslinger, CavityExp2, CavityExp3, Mekhov, Ritsch1} and the resulting cavity optomechanics 
with ultra cold atoms. The presence of an atomic ensemble in the form of a Bose Einstein Condensate (BEC) in 
such optical cavity allows a strong opto-mechanical coupling between the collective mode of the condensate with photon field. Consequently the quantum many body state of the atom can be probed by analyzing the cavity transmission. The coupled atom-photon dynamics, resulting back action, cavity induced bistability, all these together can lead to 
a number of interesting phenomena that includes self-organization of the atomic many body states \cite{SelfOrganization1, SelfOrganization, SelfOrganization2, SelfOrganization3}, bistability induced quantum phase transition \cite{Meystre} etc.  

In this context,  the deliberated  quantum optics with SOC BEC in a high finesse 
Fabry-P\'erot cavity that forms the subject matter of the current work, 
is interesting on more than one account. 
Firstly, the cavity atom interaction provides a dynamic optical lattice potential \cite{Maschler}
for the SOC Bose gas 
where the optical lattice potential is dynamically altered through its interaction with the ultra cold atomic condensate inside. This allows one to realize certain variants of extended Bose Hubbard model (eBHM). Thus far, following the seminal work of on Super-fluid (SF)-Mott-Insulator(MI) transition in ultra cold atoms \cite{Jaksch, Greiner}, such eBHM was mostly studied in the presence of prototype classical optical lattice potential. However now the dynamical nature 
of photon field contributes additional feature and profoundly influences the resulting phase diagram.  

It was already shown 
in the recent literature \cite{Trivedi, Sengupta, Radic, Cai} that a number of intriguing quantum magnetic phases can be realized by such ultra cold SOC Bose-Einstein systems in a classical optical lattice potential.  Our study of such SOC BEC inside a cavity clearly analyses such magnetic orders when 
the photon field is treated dynamically and clearly demonstrate how such magnetic phases can be detected by analyzing the transmission of photons from the cavity. As we point out, this provides an alternative way of detecting quantum magnetic phases of ultra cold atoms. Cavity spectrum has also been used to detect various other properties of the cold atomic systems such as MI-SF transition \cite{Mekhov}, detection of Landau levels in fermionic systems \cite{Ours}, phase diagram of two-component bose gas \cite{ZhangEPJD} and many more \cite{RitschRMP}. It was also proposed to create a synthetic Spin Orbit interaction in a ring cavity system \cite{ring}.

The spin orbit coupling also realizes 
a synthetic non-Abelian gauge field for such ultra cold atomic system \cite{Estienne, Spielman1} and 
consequently a spin-1/2 Bose system is also realized (in the entire work 'spin' is sometimes used in place of 'pseudo-spin'), which is fundamentally prevented by the spin-statistics theorem \cite{reviewSOC, book}. Our theoretical framework allows us to study the 
the single atom spectrum of such esoteric quantum system in the environment of a dynamical optical lattice induced by the cavity and brings out the intriguing properties of the resulting band structure. 

We unfurl the sequence of subsequent discussions as follows. 
The SOC Bose system we consider here is motivated 
by the recent experiment by NIST group \cite{Spielman1}. In section  \ref{sec:Hamiltonian} we begin with by introducing the fully second quantized Hamiltonian of such systems inside a single mode optical cavity in terms of annihilation and creation operators of photons and atoms.
 The Hamiltonian and the resulting Heisenberg Equation of motions of the field operators clearly demonstrates the dynamical nature of the optical lattice. Adiabatically eliminating the exited states of the atomic condensate we obtain an effective Hamiltonian for pseudo-spin-1/2 Bose-Einstein systems where the pseudo-spin degrees of freedom corresponds to the two lowest hyperfine states of the original multiplet of the ultra cold atomic system considered.
In the subsequent discussion, using a tight binding approximation we  derive the eBHM for the resulting system. We show that this  can be mapped suitably to the Bose-Hubbard Model of SOC Bose Gas in a classical optical lattice created due to the standing waves 
of counterpropagating laser beams \cite{Trivedi}. But now the lattice parameters being controlled by the cavity 
parameters as well as atom-photon interaction.

We arrive at our final Hamiltonian (eq. \eqref{eq:TB}) in section \ref{sec:EffectiveModel} which is an eBHM. In the subsequent section \ref{Spectrum} we study the energy spectrum of this eBHM  in the limit when atom-atom interaction vanishes. In the presence of optical lattice and synthetic non Abelian gauge field created by the spin-orbit coupling, the system shows highly intriguing band structure that features the existence of Dirac points in such bosonic system like their fermionic counterpart, a property which underscores the  spin-1/2 of such bosonic system. Then in section \ref{sec:Magnetic Phases} we discuss the various magnetic phases stabilized by the ground state of this Hamiltonian. We consider such magnetic phases in deep optical lattice regime where the orbital part is always a Mott Insulator state and the spinorial part can realize various magnetic phases 
through its texturing.

In the next section \ref{sec:CavitySpectrum} we study the probing method, i.e. how to detect various magnetic phases in an MI type of ground state through the cavity transmission spectrum. Our suggestion provides an alternative way of detecting Quantum magnetism in the ultra cold atomic systems. The role of cavity induced 
bistability in detection of such magnetic phases and the related phase transition are also discussed. We finally discuss the possibility of experimental realization of our scheme and conclude.

\section{The Model}
\label{sec:Hamiltonian}

\begin{figure}
\centering
\includegraphics[width=0.85\columnwidth , height= 
0.65\columnwidth]{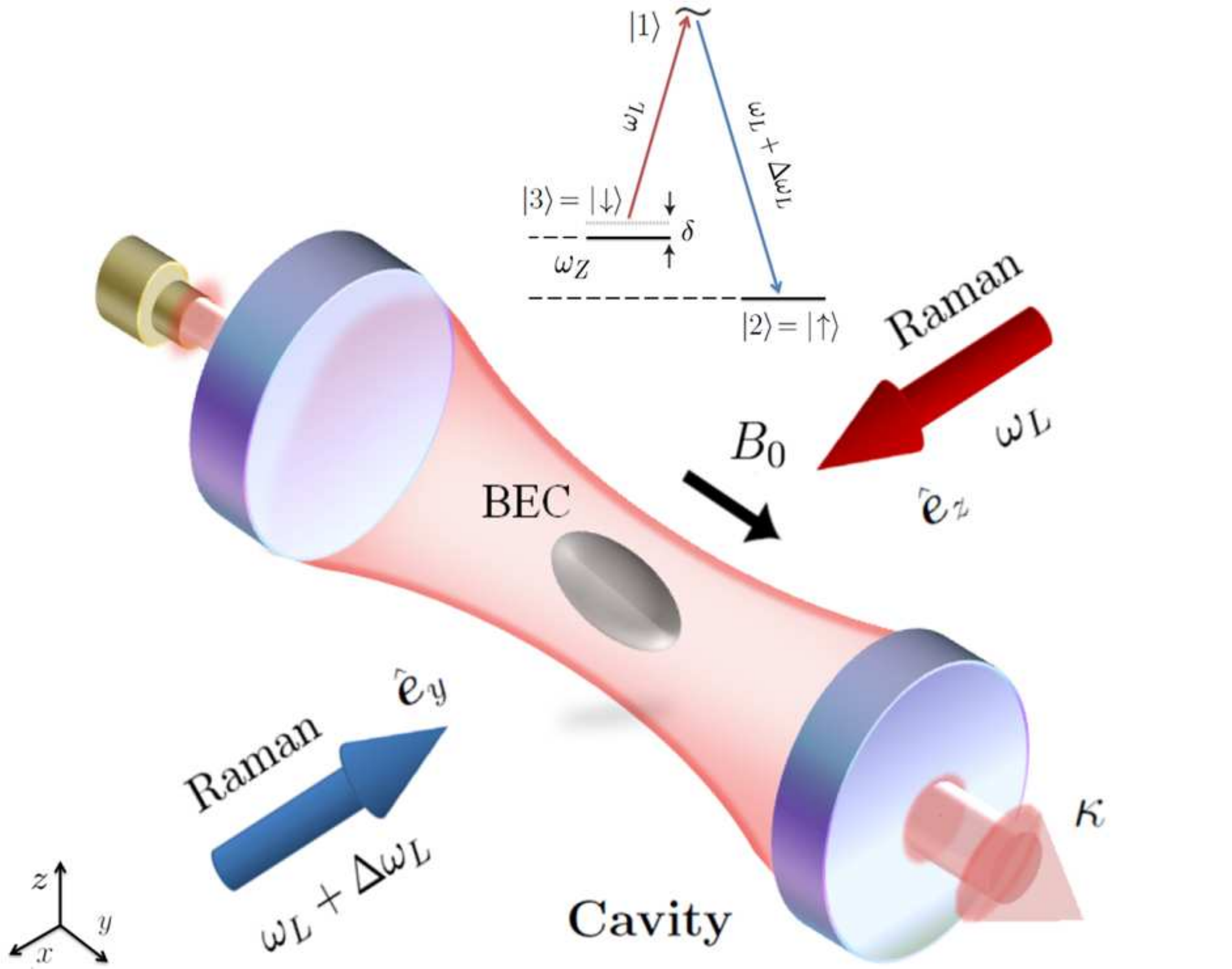} 
\caption{\textsuperscript{87}Rb BEC inside an optical cavity: SOC is created by 
two counter-propagating Raman lasers with frequencies $\omega_L$ and $\omega_L + \Delta \omega_L$ that are applied along $\hat{x}$. The Raman beams are polarized along $\hat{z}$ and $\hat{y}$ (gravity is along -$\hat{z}$). A bias field $B_0$ is applied along $ \hat{y}$ to generate the Zeeman shift. (Inset) Level diagram of the \textsuperscript{87}Rb atom. Internal states are denoted as $\ket{1}, \ket{2}, \ket{3}$. The coupling of these states is shown schematically.}
\label{fig:schematic}
\end{figure}

We consider  a condensate of $N_0$ \textsuperscript{87}Rb atoms in two internal 
states, $\ket{m_F}=\ket{1}, \ket{0}$, available in the $F=1$  manifold of 
5S\textsuperscript{1/2} electronic level. These two states are coupled by a pair of suitably detuned Raman lasers and a combination of Rashba and Dresselhaus spin orbit coupling is realized \cite{Spielman1}. This SOC BEC is now coherently driven into a linear cavity by a strong far-off resonant pump laser where it interacts with a single mode of the cavity. We consider a high Q cavity ( i.e. a cavity in which a photon takes a large number of round trips before it leaks out ) with a strong atom-field coupling. These two considerations not only enhance the atom-photon dipole interaction, but also the backaction of the atoms on the light becomes significant \cite{CavityExp1, Esslinger, CavityExp2}. The resulting atom-cavity interaction thus generates a 2D square optical lattice potential which is now dynamical \cite{RitschRMP, Maschler}.  

\subsection{The Single Particle Hamiltonian}

We derive the single-particle Hamiltonian for a two component BEC interacting with a strong, classical pump field and a weak, quantized probe field. Assuming dipole-like interaction and using rotating wave approximation we can describe a 
single atom of this system by the Jaynes-Cummings like Hamiltonian \cite{JCM}
\beq
\hh = \hh_A + \hh_C + \hh_I .
\eeq
Denoting the atomic transition frequencies as $\omega_{ij}$ and the transition operator as $\hs_{ij} = \ket{i}\bra{j}$, we express the atomic ($\hh_A$), cavity ($\hh_C$) and atom-cavity interaction ($\hh_I$) Hamiltonians as
\bse
\begin{align}
\hat{H}_A &= \frac{\hat{\bs \Pi}^2}{2m} + \hbar \omega_{12} \hs_{11} + \hbar 
\omega_{13} \hs_{11} , \\
\hat{H}_C &= \hbar \omega_c \hat{a}^\dag \hat{a} - i \hbar \eta \big ( \hat{a} 
e^{i \omega_p t} -\hat{a}^\dag e^{-i \omega_p t} \big ) , \\
\hat{H}_I &= - i \hbar g(\bs x) \big ( \hs_{12} \hat{a} - \hs_{21} 
\hat{a}^\dag + \hs_{13} \hat{a} - \hs_{31} \hat{a}^\dag \big ) .
\end{align}
\label{eq:TimeDep}
\ese
Here $\hat{\bs \Pi}^2/ 2m =(\bs p + m \bs A)^2/2m$ is the covariant momentum of the bosons. The synthetic vector potential $\bs A$ is taken to be of the form of 
$\bs A_{U(1)}+A_{SU(2)}$, where the Abelian field is \cite{SpielmanNat09} $A_{U(1)} = (0 , B_0 x , 0)$  and the spin-orbit coupling induced non-Abelian field 
is $\bs A_{SU(2)} = (\alpha \sigma_y, \beta \sigma_x, 0) $ which is \cite{Spielman1} a combination of Rashba and Dresselhouse \cite{RD-SOC} type spin-orbit coupling. When $\beta = -\alpha$ the spin orbit coupling is purely of Rashba type. Here $\alpha , \beta$ actually denote the dimensionless SOC strength in the unit of $\frac{\hbar K}{\pi m}$, where $K$ is the wave number corresponding to the cavity photon.
 $\hat{\sigma}_{x, y,z}$ are $2 \times 2$ spin-1/2 representation of Pauli matrices. $\eta$ is the coupling between the pump and the cavity, $\omega_p$ is the frequency of the pump laser which we set to be $\omega_L + \Delta \omega_L$, $\omega_c$ is the frequency of the cavity photon which is almost in resonance with the pump beam, $\Delta_c = \omega_p - \omega_c = \Delta \omega_L \approx \kappa$, with $2\kappa$ being the cavity decay line-width. The operator $\hat{a}$ ($\hat{a}^\dag$) annihilates (creates) one cavity photon. 

$g^2(\bs x)$ is the cavity mode function which varies as the spatial mode profile and we take $g^2(\bs x) = g_0 [\cos^2(Kx) + \cos^2(Ky)]$, where $g_0$ is the coupling strength of the atom and cavity field. We also assume the wave vector along $x$ and $y$ directions are same, namely $K_{x}=K_{y}=K$.
For simplicity we assume both the transitions $\ket{2} \leftrightarrow \ket{1}$ and $\ket{3} \leftrightarrow \ket{1}$ have the same coupling with the cavity. Assuming the atoms to be in the same motional quantum-state, the coupling $g_0$ is assumed to be identical for all atoms. In order to remove the time-dependence of the above Hamiltonian we perform a unitary transformation on the above Hamiltonian with $\hat{U}(t) = \exp [i \omega_p t \big ( \hs_{11} + \ha^\dag \ha \big) ]$. Using the Baker's lemma the following Hamiltonians are obtained (see appendix \ref{app:Baker}): 
\bse
\begin{align}
\hh_A &= \frac{\hat{\bs \Pi}^2}{2m} - \hbar \Delta^a_{12} \hs_{11} - \hbar 
\Delta^a_{13} \hs_{11} , \\
\hh_C &= - \hbar \Delta_c \ha^\dag \ha - i \hbar \eta \big (\ha -\ha^\dag \big ) + \kappa \ha^\dag \ha
, \\
\hh_I &= - i \hbar g(\bs x) \big ( \hs_{12} \ha - \hs_{21} \ha^\dag + \hs_{13} 
\ha - \hs_{31} \ha^\dag \big ) .
\end{align}
\label{eq:TimeInd}
\ese
The atom-pump detuning is denoted as $\Delta_{ij}^a = \omega_p - \omega_{ij}$. From now we denote $\Delta_a = \Delta_{12}^a + \Delta_{13}^a$. The extra term $\kappa \ha^\dag \ha$ appearing in $\hh_C$ can be justified in the following way: in the presence of external pumping of atoms the system becomes an open quantum system and hence dissipation effects must be incorporated. This is done using the master equation approach for (atom-field) density matrices \cite{Mekhov, Zoller}. Thus the effect of photon loss due to cavity decay line-width ($\kappa$) gets incorporated.

\subsection{The Many-Body Hamiltonian}
Following references \cite{Mekhov, Meystre} we now derive the full many-body Hamiltonian for this system. For that we construct a matrix of all the transition operators and project it onto the full many-body space. This causes the transition operator $\hs_{ij}$ to pick up the product of $\hp^\dag_i$ and $\hp_j$. So the final form of the many-body Hamiltonian becomes
\bea
\mH_A &=& \inx{ \hp^\dag_2 (\bs x) \Big ( \frac{\hat{\bs \Pi}^2}{2m} \Big ) 
\hp_2 (\bs x) + \hp^\dag_3 (\bs x) \Big ( \frac{\hat{\bs \Pi}^2}{2m} \Big ) 
\hp_3 (\bs x) \nn \\
&& + \hp^\dag_1 (\bs x) \Big ( \frac{\hat{\Pi}^2}{2m} - \hbar \Delta_a \Big ) 
\hp_1 (\bs x)} ,
\eea
Here $\hp_{i}(\bs x), \hp^{\dag}_{i}(\bs x)$ are the annihilation and creation operators for atom at position $\bs x$ in the spin-state  $\ket{i}$. They obey usual bosonic commutation relations
\bse
\begin{align}
\Big [ \hp^{\pdag}_i (\bs x), \hp^\dag_j (\bs x') \Big ] &= \delta^3(\bs x - \bs 
x') \delta_{ij} \\
\Big [ \hp_i (\bs x), \hp_j (\bs x') \Big ] &= \Big [ \hp^\dag_i (\bs x), 
\hp^\dag_j (\bs x') \Big ] =0 .
\end{align}
\ese
Since the cavity field operators commute with the atomic operators the 
Hamiltonian $\mH_C$ remains unchanged in the second-quantized notation.  In our 
analysis we assume the pump mode is so chosen that its interaction with the 
atoms is solely along the $\hat{z}$ axis, allowing us to exclude its dynamics on 
$x-y$ plane. The two body interaction between the atoms in same and different spin state is modelled through \cite{Spielman1},
\bea
\mH_U = \frac{U}{2} \inx{\dsi{2} \dsi{2} \si{2} \si{2} + \dsi{3} \dsi{3} \si{3} 
\si{3} \nn \\
+ \lambda \dsi{2} \dsi{3} \si{2} \si{3} }, 
\eea 
where the intra-species interaction strength is measured by $U = 4\pi a_s^2 
\hbar^2/m$ and the inter-species interaction is measured by $\lambda U$, where the parameter $\lambda$ is decided by the laser configuration. Here $a_s$ is s-wave scattering length. Next, the many-body interaction between the atom and cavity can be modeled as 
\beq
\mH_I = -i \hbar \inx{ \dsi{1} \ha \si{2} + \dsi{1} \ha \si{3} + \text{h.c.} 
}g(\bs x) .
\eeq
Now we calculate the Heisenberg equations of evolution for various field 
operators (say $\hat{A}$), $ i \hbar \partial_t \hat{A} = [\hat{A},\mH]$ :
\bse
\begin{align}
\evolve{\si{1}} &= - i \Big ( \frac{\hat{\bs \Pi}^2}{2\hbar m} - \Delta_a \Big ) 
\si{1} - g(\bs x) \ha \big ( \si{2} + \si{3} \big ), \label{evoa} \\
\evolve{\si{2}} &= - i \Big ( \frac{\hat{\bs \Pi}^2}{2\hbar m} + \frac{U}{\hbar} 
\dsi{2}\si{2} + \frac{U \lambda }{\hbar} \dsi{3}\si{3} \Big ) \si{2} + g(\bs x) 
\ha^\dag \si{1} , \label{evob} \\
\evolve{\si{3}} &= - i \Big ( \frac{\hat{\bs \Pi}^2}{2\hbar m} + \frac{U}{\hbar} 
\dsi{3}\si{3} + \frac{U \lambda}{\hbar} \dsi{2}\si{2} \Big ) \si{3} + g(\bs x) 
\ha^\dag \si{1} , \label{evoc} \\
\evolve{\ha(t)} &= i \Delta_c \hat{a}(t) + \eta + \inx{\dsi{2} g(\bs x) \si{1} + 
\dsi{3} g(\bs x) \si{1}}. \label{evod}
\end{align}
\label{evolve}
\ese
In the evolution of atomic operators the first term describes the free evolution 
of the atomic states. In \eqref{evoa} the second term describes the 
absorption of cavity photon by an atom, causing an excitation from $\ket{2}$ or 
$\ket{3}$ to the excited state $\ket{1}$. Similarly in \eqref{evob} or \eqref{evoc} the second term describes the emission of a cavity photon 
followed by the relaxation of an atom from state $\ket{1}$ to $\ket{2}$ or 
$\ket{3}$. The first term in \eqref{evod} is the free 
evolution term and the last two terms are the two additional driving terms 
of the field, one by the pump and the other by the emission of an atom due to 
relaxation from state $\ket{1}$ to $\ket{2}$ or $\ket{3}$. 

In order to preserve the BEC in its ground state we must avoid heating, 
primarily caused by spontaneous emission from the atoms. The excited state vary 
with a time scale of $1/\gamma$ (atomic line-width) and the ground state and 
cavity photons evolve with a time scale of $1/\Delta_a$. Hence by choosing a 
large atom-pump detuning, $\Delta^a_{ij} \gg \gamma$ we can adiabatically 
eliminate the excited states from the dynamics of our system \cite{Mekhov}. By setting 
$\partial_t \si{1} = 0$ we obtain:
\beq
\si{1} = - \frac{i}{\Delta_a} g(\bs x) \ha (t) \big [ \si{2} + \si{3} \big ]. 
\eeq
Inserting this into \eqref{evolve} we get
\bse
\begin{align}
\evolve{\si{2}} &= -i \Big [ \frac{\hat{\bs \Pi}^2}{2 \hbar m} + \frac{U}{\hbar} 
\dsi{2}\si{2} + \frac{U \lambda }{\hbar} \dsi{3}\si{3} \nn \\
& \qquad \qquad \qquad + \frac{g^2(\bs x)}{\Delta_a} \ha^\dag \ha \Big ] \si{2} 
- i \frac{g^2(\bs x)}{\Delta_a} \ha^\dag \ha \si{3}, \\
\evolve{\si{3}} &= -i \Big [ \frac{\hat{\bs \Pi}^2}{2 \hbar m} + \frac{U}{\hbar} 
\dsi{2}\si{2} + \frac{U \lambda }{\hbar} \dsi{3}\si{3} \nn \\
& \qquad \qquad \qquad + \frac{g^2(\bs x)}{\Delta_a} \ha^\dag \ha \Big ] \si{3} 
- i \frac{g^2(\bs x)}{\Delta_a} \ha^\dag \ha \si{2}, \\
\evolve{a(t)} &= i \Big [ \Delta_c - \frac{1}{\Delta_a} \int d \bs x g^2(\bs x) 
\Big [ \dsi{2} \si{2} + \dsi{3} \si{3} \nn \\ 
& \qquad \qquad \qquad + \dsi{2} \si{3} + \dsi{3} \si{2} \Big ] \ha + \eta .
\end{align}
\label{effective}
\ese
This set of equations is a characteristic of cavity opto-mechanical system \cite{StamperKurnBook}. Here we have developed them specifically for a SOC-BEC system. Since we have adiabatically eliminated the excited state $\ket{1}$ from the dynamics, from now onwards we drop the notation of $\{2,3\}$, and use $\{\upa, \dna\}$ 
instead to use the language of 'pseudo-spins'. In other words, the two laser-dressed hyperfine states $\ket{F=1, m_F= 0}$ and $\ket{F=1, m_F= 1}$ of the \textsuperscript{87}Rb atoms are now mapped to a synthetic spin-1/2 system (hence pseudo-spin), with states labeled as $\ket{\upa}$ and $\ket{\dna}$. It must be noted that there exists no real spin-1/2 bosonic systems in nature due to spin-statistics theorem, but with the help of lasers we could realize such a system in ultra cold atomic condesnate  \cite{Spielman1}. In further sections we will show this strange property of the system leads to some interesting (for bosonic systems) results which are unconventional 
in bosonic systems. 

Now the dynamics of the atoms effectively comprises of the dynamics of a two species (denoted by their pseudo-spin label) bosons coupled by spin-orbit interaction. The effective Hamiltonian $\mH_{eff}$ which captures the effective dynamics of the system described in \eqref{effective}, $i \hbar \partial_t \hp_{\upa,\dna}(\bs x) = 
[\hp_{\upa,\dna} (\bs x), \mH_{eff}]$ and $i \hbar \partial_t \ha = [\ha, 
\mH_{eff}]$.
\bea
\mH^{(1)}_{eff} = \int d \bs x \bs \hp^\dag(\bs x) \Big ( \frac{\hat{\bs 
\Pi}^2}{2m} + U_{lat}  \Big ) \bs \hp(\bs x) + \hat{H}_c
\nn \\
+ \frac{1}{2} \int d \bs x \sum_{s,s'} U_{s,s'} \dsi{s} \dsi{s'} \si{s'} \si{s} ,
\label{eq:Heff}
\eea
Here $s,s' \in \{ \upa, \dna\}$. For simplification of notations we have defined a column vector $\bs \hp = (\hp_{\upa}, \hp_{\dna})^T$. The atom-atom interaction strength is denoted as $U_{\upa, \upa} = U_{\upa, \upa} = U$ and $U_{\upa, \dna} = U_{\dna , \upa} = \lambda U $. One can note the atom-cavity coupling has lead to the formation of an optical lattice \cite{Mekhov}, which is $U_{lat} = V_0 [\cos^2(K x) + \cos^2 (K y)]$. Here $V_0$ is the depth of the well, $V_0 = \hbar U_0 \ha^\dag \ha $ and $U_0 =g_0^2/\Delta_a$ is the effective atom-photon coupling strength. Now since the lattice depth has become a (photon number) operator, it is no longer a classical lattice but a quantum lattice. In our calculations we have taken an Nd:Yag (green) laser source of $\lambda = $1064 nm (hence the lattice constant is $a_0 = \lambda /2 =$ 532nm). The kinetic energy of an atom carrying one unit of photon momentum, $|\bs p| = \hbar K$ describes the characteristic frequency of the center of mass motion of the cloud. Thus the relevant energy scale is $E_r = \hbar^2 K^2/2m$ (recoil energy), in the units of which we measure all other energies involved in the problem. For our case the lattice recoil frequency is $\omega_r = E_r/ \hbar = 12.26$ kHz.

\subsection{The Extended Bose-Hubbard Model}
To investigate various interesting phases of this system through the cavity spectrum,  first we establish an equivalence of the effective Hamiltonian obtained in \eqref{eq:Heff} in a cavity induced quantum optical lattice 
with a prototype Bose-Hubbard model
in a classical optical lattice. Using  tight binding approximation this is done as follows. By constructing maximally localized eigenfunctions at each site of the lattice we expand each component of the atomic field operator $\hp_s$ in the basis of Wannier functions \cite{Kittel}, 
\beq
\hp_s (\bs r) = \sum_{i} \hat{b}_{s i} w(\bs r - \bs r_i ),
\label{eq:WanBasis}
\eeq
$\hat{b}^\dag_{s i}$ is a bosonic operator that creates an atom in pseudo-spin state $| s \rangle$ ($s = \{ \upa, \dna \}$) at site $i$ of the optical lattice.  However, in presence of a gauge potential the Wannier functions pick up a gauge dependent phase and should be modified as 
\beq
w(\bs r - \bs r_i) \rightarrow W(\bs r - \bs r_i) = e^{-i \frac{ m}{\hbar}\int_{\bs r_i}^{\bs r} \bs A(\bs r') \cdot d\bs l}w(\bs r - \bs r_i).
\label{Wann}
\eeq

First we show that under nearest neighbor approximation (i.e. hopping is permitted in between two adjacent sites only), the gauge transformed Wannier function in \eqref{Wann} forms a valid basis for the Hilbert space and then we expand the effective Hamiltonian in \eqref{eq:Heff} in this basis. We denote $w(\bs r - \bs r_i)$ as $w_i(\bs r)$. The norm of the gauge transformed Wannier functions becomes equal to unity since the gauge transformation only introduces a phase factor. So we check for orthogonality only. The inner product is 
\begin{widetext}
\bea
\int d \bs r W^*_i (\bs r) W_j (\bs r) &=& \int d\bs r e^{-i \Big [ \alpha  
\sigma_y (x_j - x_i) - \alpha \sigma_x (y_j - y_i)  + B_0 x(y_j - y_i)  \Big ]} 
w^*_i (\bs r) w_j (\bs r)  \nn \\
&=& e^{- i \Big [ \alpha  \sigma_y (x_j - x_i) - \alpha \sigma_x (y_j - y_i)  
\Big ]} \int dx e^{-i B_0 x(y_j - y_i)} w_i^*(x)w_j(x) \int dy w_i^*(y)w_j(y).\nn \\
& & \label{orthogonality}
\eea
\end{widetext}
For integration along x-axis, $y_j-y_i =0$ the first integral in 
\eqref{orthogonality} causes the entire express to vanish to zero, owing to the 
orthogonality of the Wannier functions $w_i(x)$, i.e. $\int d \bs r w^*_i(\bs r) w_j(\bs r) = \delta_{ij}$. For integration along y-axis 
second integral in \eqref{orthogonality} makes the total integral zero because 
of the orthogonality of the Wannier functions $w_i(y)$. Hence we establish 
orthonormality, under nearest-neighbor approximation :
\beq
\int d \bs r W^*_i (\bs r) W_j (\bs r) = \delta_{ij}.
\eeq
The action of the covariant derivative on this modified Wannier function can be 
shown to be (recall $\hat{\bs \Pi} = -i \hbar \bs \nabla  + m \bs A$)
\beq 
\hat{\bs \Pi} W_i(\bs r) = e^{-i \frac{m}{\hbar} \int_{\bs r_i}^{\bs r_j} \frac{\hbar}{i} \bs A(\bs r') \cdot d\bs l} \bs \nabla w_i(\bs r). 
\label{covar} 
\eeq
Substituting Eq. \eqref{eq:WanBasis}  in the  effective Hamiltonian in \eqref{eq:Heff} and using Eqs. (\ref{orthogonality}) and 
(\ref{covar}) we obtain
\bea
\hh_A &=& \int d^2r \bs \hp^\dag (\bs r) \frac{\hat{\bs \Pi}^2}{2} \bs \hp (\bs 
r) = \frac{1}{2} \sum_{i,j} \bem \hat{b}^\dag_{\uparrow i} & 
\hat{b}^\dag_{\downarrow i} \eem
\int d^2r W^*_i(\bs r ) \bs \Pi^2 W_j(\bs r ) \bem \hat{b}_{\uparrow j} \\ 
\hat{b}_{\downarrow j} \eem \nn \\
&=& \sum_s \Big ( \sum_i E_{ii} \hat{b}^\dag_{s i} \hat{b}_{s i} + \sum_{<i,j>} 
\hat{b}^\dag_{s i} E_{ij}e^{-i \phi_{ij}} \hat{b}_{s j} \Big ) = E_0 \hN + E \hB 
.
\\
\hh_I &=& \int d^2 r \sum_{ s }\hp^\dag_s (\bs r) \hat{U}_{lat} \hp_s (\bs r)  \nn \\
&=& 
U_0 \ha^\dag \ha \sum_{i,j} \bem b^\dag_{\uparrow i} & b^\dag_{\downarrow i} 
\eem
\int d^2r W^*_i(\bs r ) [\cos^2 (Kx) + \cos^2 (Ky)] W_j(\bs r ) \bem 
\hat{b}_{\uparrow j} \\ \hat{b}_{\downarrow j} \eem \nn \\
&=& U_0 \ha^\dag \ha  \sum_s \Big ( \sum_i J_{ii} \hat{b}^\dag_{s i} \hat{b}_{s 
i} + \sum_{<i,j>} \hat{b}^\dag_{s i} J_{ij} e^{-i \phi_{ij}} \hat{b}_{s j}  \Big 
) = \hat{U}_0 \ha^\dag \ha (J_0 \hN + J_1 \hB).
\eea
Unlike the case of the BH model in 
a classical optical lattice \cite{Jaksch}, for a lattice generated by quantum 
light we have treated the matrix elements of the potential and kinetic energy 
separately. It is because of the presence of the term $\ha^\dag \ha$ in the potential term. So the extended BH Hamiltonian becomes 
\bea
\mH^{(2)}_{eff} =& E_0 \hN + E_1 \hB + \hbar U_0 \ha^\dag \ha (J_0 \hN +J_1 \hB) 
- \hbar \Delta_c \ha^\dag \ha
\nn \\
& - i \hbar \eta (\ha - \ha^\dag) + \frac{1}{2} \sum_{i, s,s'} U_{s,s'} 
b^\dag_{is} b^\dag_{is'} b^{\pdag}_{is'} b^{\pdag}_{is} ,
\label{eq:Hubbard}
\eea
Here $E_0$ ($E_1$) and $J_0$ ($J_1$) are the on-site (off-site) elements of $E_{ij}$ and $J_{ij}$, respectively and these are :
\bse
\begin{align}
E_{ij} &= \frac{\hbar^2}{2m} \int d^2r w^*_i(\bs r ) \bs \nabla^2 w_j(\bs r ), \\
J_{ij} &= \int d^2r w^*_i(\bs r ) 
[\cos^2(Kx) + \cos^2(Ky)] w_j(\bs r ) .
\end{align}
\ese
$\hN = \sum_{s , i} \hat{b}^\dag_{s i} \hat{b}_{s i}$ is the total atom number operator and $\hB = \sum_s \sum_{<i,j>} \hat{b}^\dag_{s i}  e^{-i \phi_{ij}}\hat{b}_{s j} $ is the nearest neighbor hopping operator, for the full form of $\hB$ see appendix \ref{app:Hopping}. Here $\phi_{ij}$ is the phase acquired by an atom while hopping from lattice site $i$ to $j$ :
\beq
\phi_{ij}= \alpha  \sigma_y (x_j - x_i) + \beta \sigma_x (y_j - y_i)  + \mathbb{1} B_0 x_i (y_j - y_i) 
\eeq 
Here $\mathbb{1}$ is a $2 \times 2$ unit matrix. Because of the dynamical nature of the lattice ( the coefficient 
term for the lattice potential involves operators) $E_{ij}$ and $J_{ij}$ are treated separately, otherwise the hopping amplitude would be identified with $t = E_1 + J_1$ and the chemical potential with $\mu = E_0 + J_0$.

\section{Elimination of Cavity Degrees of Freedom}

\subsection{The Effective Model}
\label{sec:EffectiveModel}

The interplay of energy scales associated with the spin orbit coupling, motion of atoms in a dynamical lattice and atom-atom interactions brings out a richer and more complex dynamics, as compared to the usual BH model \cite{Jaksch, Mekhov}, which we try to capture through the light coming out of the cavity.  To facilitate  further discussion on dynamics governed by  \eqref{eq:Hubbard}  we shall do certain simplifications based on the typical 
experimetal systems. Following typical experimental situation \cite{Esslinger, CavityExp1, CavityExp2} we work under bad cavity limit where we assume the cavity field reaches its stationary state very quickly than the time scale involved with atomic dynamics. 
Hence it is reasonable (at least for $t > 1/\kappa ,$) to replace the light 
field operators with their steady state values, and thus adiabatically eliminate 
the cavity degrees of freedom from the Hamiltonian \eqref{eq:Hubbard} so that it 
depends only on the atomic variables. It will be useful to remember this process is distinct from the adiabatic elimination of the excited state $\ket{1}$, carried in the previous section. The evolution of light field operators can be obtained from \eqref{eq:Hubbard} 
as 
\bea  \partial_t \ha  = \frac{1}{i \hbar}[\ha, \hh_{eff}^{(2)}]  = - \hat{D} 
\ha + \eta, \label{eq:Photon}
\eea
where $\hat{D} =   \kappa + i [ U_0 (J_0 \hN + J_1 \hB) - \Delta_c ]$ is a complex operator. Assuming the total number of atoms to be fixed we can 
replace the atom number operator by a fixed quantity $N_0 = \langle \hN \rangle 
$, and due to the presence of atoms an effective detuning is obtained as 
$\Delta_c' = \Delta_c - U_0J_0 N_0$. Setting $\partial_t \ha = 0$ we get the 
steady state value $\ha^{(s)} = \eta/ \hat{D}$ and then expand $\hat{a}$ with 
respect to the hopping matrix $\hB$ :
\beq 
\ha^{(s)} \approx \frac{\eta}{\kappa - i \Delta_c'} \Big [ 1 - \frac{i U_0 
J_1}{\kappa - i \Delta_c'} \hB - \frac{U_0^2 J_1^2}{(\kappa - i \Delta_c')^2} 
\hB^2 + ... \Big ]  \eeq
Substituting this in the Hamiltonian \eqref{eq:Hubbard} we obtain the effective 
Hamiltonian, expressed in terms of atomic variables :
\bea
\mH^{(3)}_{eff} = - \mj \hB + \mf \hB^2 + ... + \frac{1}{2} \sum_{i, s,s'} 
U_{s,s'} \hb^\dag_{is} \hb^\dag_{is'} \hb^{\pdag}_{is'} \hb^{\pdag}_{is}.
\label{eq:cavity-Hubbard}
\eea
\bse
\begin{align}
\mj / J_1 &= U_0 \eta^2 \frac{\kappa^2 - \Delta_c'^2}{(\kappa^2 + \Delta_c'^2)^2} 
- E/ J_1, \\
\mf/ J_1^2 &= 3 U_0^2 \eta^2 \Delta_c' \frac{3\kappa^2 - \Delta_c'^2}{(\kappa^2 + \Delta_c'^2)^4} .
\end{align}
\ese
\begin{figure}
\centering
\includegraphics[width= 0.8 \columnwidth, height= 0.65 
\columnwidth]{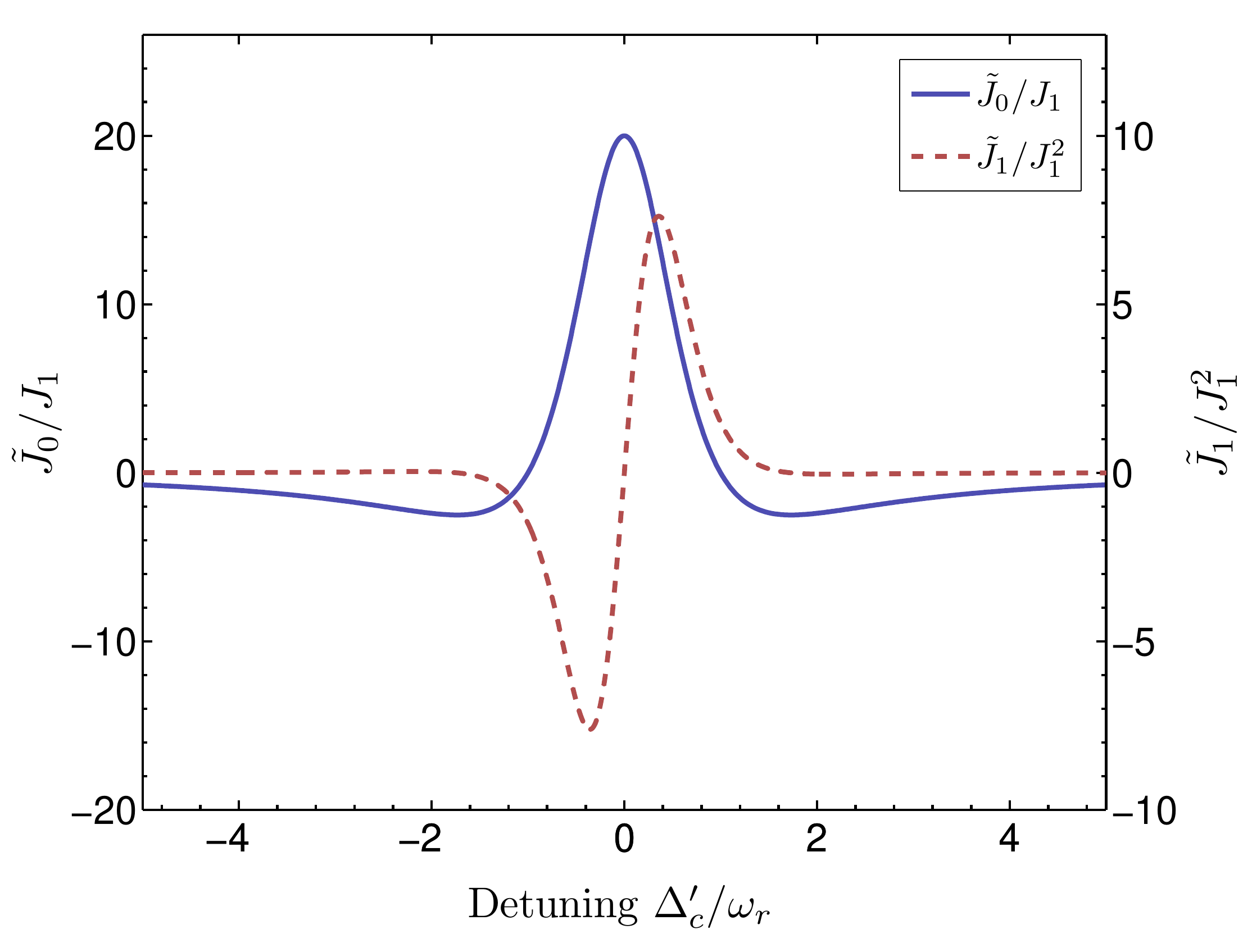}
\caption{Variation of the two coefficients used in \eqref{eq:cavity-Hubbard} 
with effective detuning. The experimental parameters are set to be \{$\eta$, 
$\kappa$, $U_0$, $J_0$\} = \{10, 1, 0.2, 2\}$\omega_r$. }
\label{fig:JF}
\end{figure}
The parameter $\mj$ is the rescaled hopping amplitude, where the scaling 
factor is introduced by the cavity parameters and that of atom-photon 
interaction strength. Its variation with cavity detuning is shown in Figure 
\ref{fig:JF}. Note $\tilde{J}_0$ can be made to vanish by setting $\Delta_c' = 
\kappa$, and similarly $\tilde{J}_1$ vanishes when $\Delta_c' = \sqrt{3}\kappa 
$.

 It is clear from (\ref{eq:cavity-Hubbard}) that cavity-atom coupling induces higher order hoppings feasible through terms like $\hB^{(n)}$.  Also the amplitude of there terms are well controllable through cavity parameters allowing to study higher order atom-atom correlations in these systems. Through suitable choice of cavity parameters, we suppress all higher order terms starting from $\hB^2$. This renders $\mH^{(3)}_{eff}$ to a tight-binding Hamiltonian 
\cite{Kittel},  which has incorporated in itself the effects of cavity, Abelian and non-Abelian gauge field altogether :
\beq
\mH^{(4)}_{eff} = - \mj \hB + \frac{1}{2} \sum_{i, s,s'} 
U_{s,s'} \hb^\dag_{is} \hb^\dag_{is'} \hb^{\pdag}_{is'} \hb^{\pdag}_{is}. 
\label{eq:TB}
\eeq 
 This is our effective Bose Hubbard  Hamiltonian, on which rest of the work is built on. The hopping amplitude is $\mj$. The hopping operator $\hB$ now contains all the information about spin orbit coupling. However it may be pointed out that 
apart from modifying bare hopping amplitude $J_0$ to the rescaled $\mj$, the 
cavity also triggers long-range correlations via higher order terms in $\hB$ 
which we ignored. In fact in presence of a dynamical lattice both the atom and 
photon operators evolve, in accordance with their corresponding (coupled) 
Heisenberg equations \cite{Mekhov}. One can solve this pair of equations 
simultaneously to study the full self-organization. However assuming the atoms 
fall through the cavity light field sufficiently faster (much before the atoms affect the cavity photon) we ignore the back action of the atoms on the cavity light \cite{StamperKurnBook}. Self-organization of atoms in the lattice \cite{SelfOrganization, SelfOrganization2} can in itself be a separate direction to pursue, facilitating the study of self-organized checkerboard phase \cite{Deng}, supersolid phase \cite{Supersolid}, or quantum spin-glass phase \cite{Spin-glass}. 

{In the following subsection we analyze the complete energy spectrucm of the  effective Hamiltonian in \eqref{eq:TB} in the non-interacting limit first. For this subsection only, we switch off inter atomic interaction, namely $U_{s,s'}  = 0$, which can be achieved through the tuning of Feschbach resonance \cite{ResoRMP}. 

\subsection{The Spectrum: non interacting limit}
\label{Spectrum}
The rescaling of the hopping amplitude by cavity parameters allows a number of 
physical properties to be controlled through such parameters. We study 
the spectrum of this tight-binding Hamiltonian obtained in \eqref{eq:TB}. We reiterate that the analysis in this 
section is in the absence of atom-atom interaction. We shall show that the resulting system yields 
 two interesting spectra namely, the Hofstadter butterfly spectrum \cite{Hofstadter} and the Dirac spectrum. The emergence of Hofstadter spectrum is natural as the considered non interacting bosonic system mimics the motion of Bloch particle (a quantum mechanical particle in a periodic lattice potential) in presence of a uniform U(1) gauge field. The energy levels of such particle is the Hofstadter spectrum- a butterfly like structure is revealed when the energy values of the Bloch particle is plotted against the Abelian Flux inserted. Such is the case in the absence of Spin Orbit coupling  ($\alpha = 0$) where the Hamiltonian in \eqref{eq:TB} becomes identical with a Harper Hamiltonian, which can be obtained through Peierl's substitution in the usual tight-binding Hamiltonian \cite{Hofstadter}. Recently, two groups at the M.I.T and in Munich have experimentally realized such butterfly spectrum in cold atomic systems \cite{Bloch-Ketterle}. However, compared to those systems, in the present case  one can control (through suitable choice of $\mj$) the energy scale of the butterfly structure just by suitably tuning the cavity parameters. The effects of non-Abelian gauge field  on such butterfly structure, was also studied \cite{Kubasiak}.

Next we show how the Dirac spectrum emerges. For this the Hamiltonian in \eqref{eq:TB} is diagonalized in appendix \ref{app:diagonalization} and the spectrum obtained is:
\bea
{E_\pm}/{\mj}= 2 \cos \alpha \cos k_x + 2 \cos \beta \cos (k_y - 2m\pi \Phi) \nn \\ \pm \sqrt{\sin^2 \alpha \sin^2k_x + \sin^2 \beta \sin^2 (k_y - 2m\pi \Phi)} ,
\label{diracspec}
\eea
where $(m,n)$ is a lattice point. The energy values are plotted against particle momentum and a Dirac like spectrum is obtained in Fig. \ref{fig:diracpoints}.  

The band-splitting in the spectrum becomes evident as 
soon as the effects of SOC is incorporated, showing a band gap ($E_g$) of
$ E_g/ \mj = 4\sin \alpha \sqrt{\sin^2k_x + \sin^2(k_y -2m\pi\Phi)} $, where the gap can be tuned by the cavity as well (through $\mj$). Also in the first Brillouin zone the band gap is maximum when $(k_x,k_y) \in \{ (\pm \pi/2 , \pm \pi/2) \}$ and $E_g^{max} /\mj= 4\sqrt{2} \sin \alpha \equiv W$. It is possible to carry out a 
bandgap measurement in such systems through Bragg spectroscopy \cite{DiracExpt}, 
through which one can measure the non-Abelian flux inserted in the system. However, the gap vanishes when both $\sin k_y = \sin k_x = 0$. In the first Brillouin Zone (by setting $\Phi = 0$) this can happen for $(k_x , k_y ) \in \{ (0,0) , (\pm \pi , 0) , (0 , \pm \pi) , (\pm \pi , \pm \pi) \} \equiv \bs k_D$. In the vicinity of these points the effective low energy behavior can be described (see appendix \ref{app:diagonalization} for details) by a Dirac like Hamiltonian, 
\beq
\hat{H}_{eff} = - \sum_{\bs p} \hp_{\bs p}^\dag \hat{H}_D \hp_{\bs p}  ,  
\hat{H}_D = c_x \gamma_x p_x + c_y \gamma_y p_y.
\label{eq:Dirac}
\eeq
Here $\hat{H}_D $ is a Dirac Hamiltonian, $\bs p = \bs k - \bs k_D$, but the field operators $\hp_{\bs p}$ are 
bosonic annihilation operators. The 
gamma matrices $\gamma_0 = \bs 1, \gamma_1 = \gamma_x = \sigma_y, \gamma_2 = \gamma_y = \sigma_x$ are the $2+1$ dimension representation of Clifford algebra, $\{ \gamma_i, \gamma_j\} = 2\delta_{ij}$. The speeds of light $c_x = 2\sin \alpha, c_y=2\sin \beta$ are now anisotropic. As shown in the Figure \ref{fig:diracpoints}, through this anisotropy the SOC strength can be used as a handle to controlling the shape of the Dirac cones. We refer the 'Dirac-like' points $\bs k_D$ in our bosonic system also as Dirac points. Near $\bs k_D$ the excitation quasi particles are mass-less bosons having a dispersion relation linear in $\bs k$, the slope of which is controlled by adjusting the spin-orbit coupling strength.
\begin{figure}
\subfloat{\includegraphics[width=0.5\columnwidth,height=0.4 
\columnwidth]{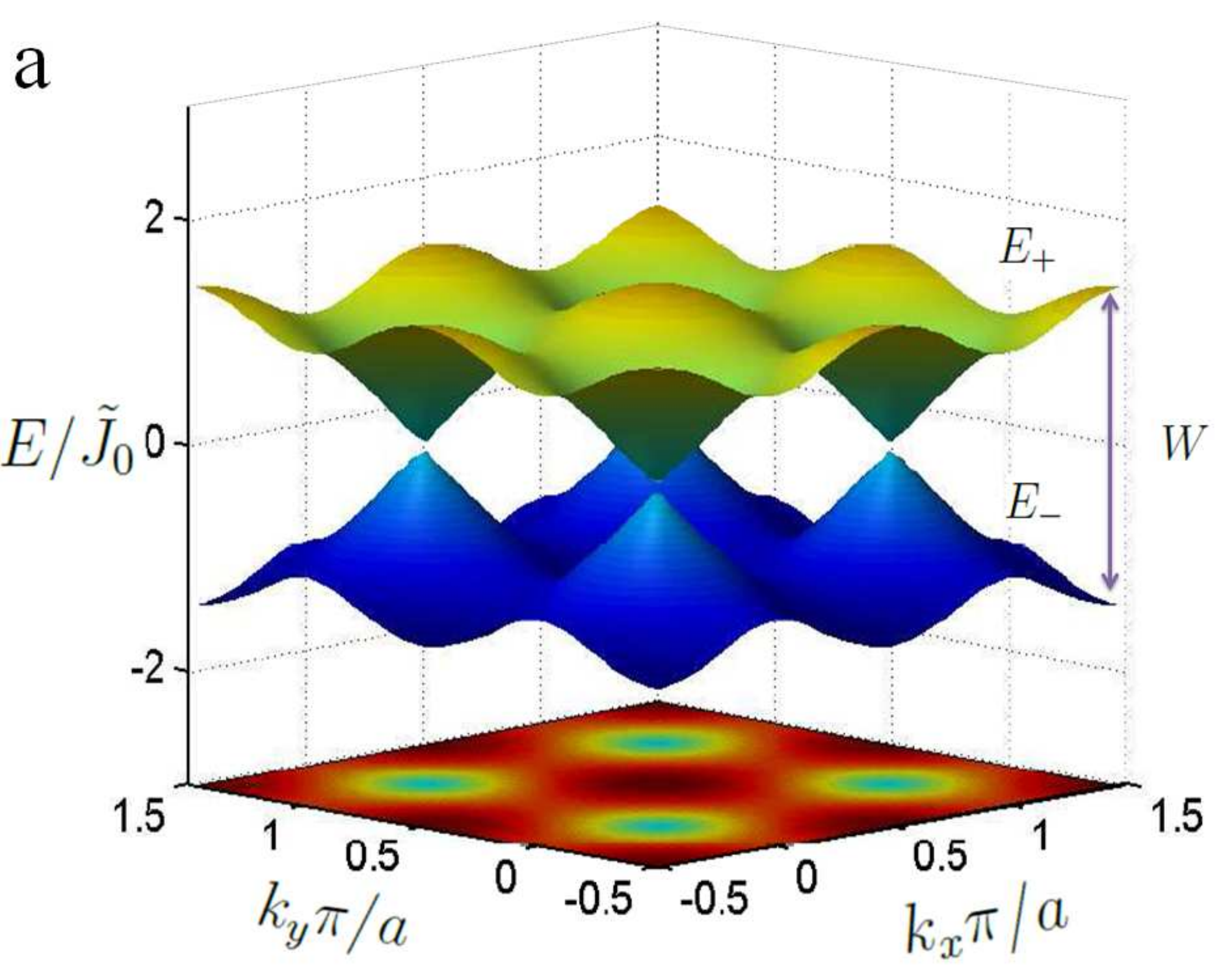} }
\subfloat{\includegraphics[width=0.5\columnwidth,height=0.4
\columnwidth]{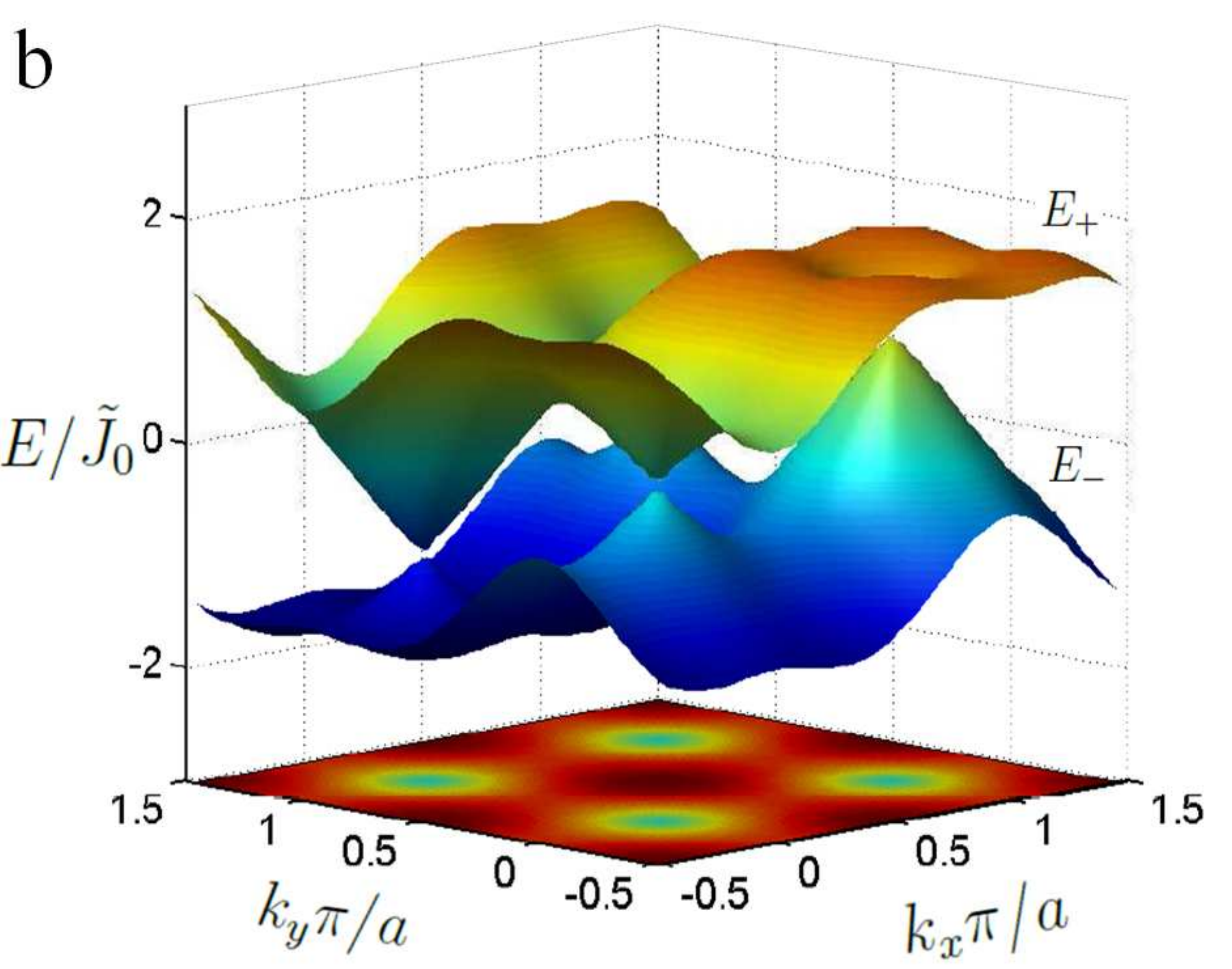} } \\
\subfloat{\includegraphics[width=0.5\columnwidth,height=0.36
\columnwidth]{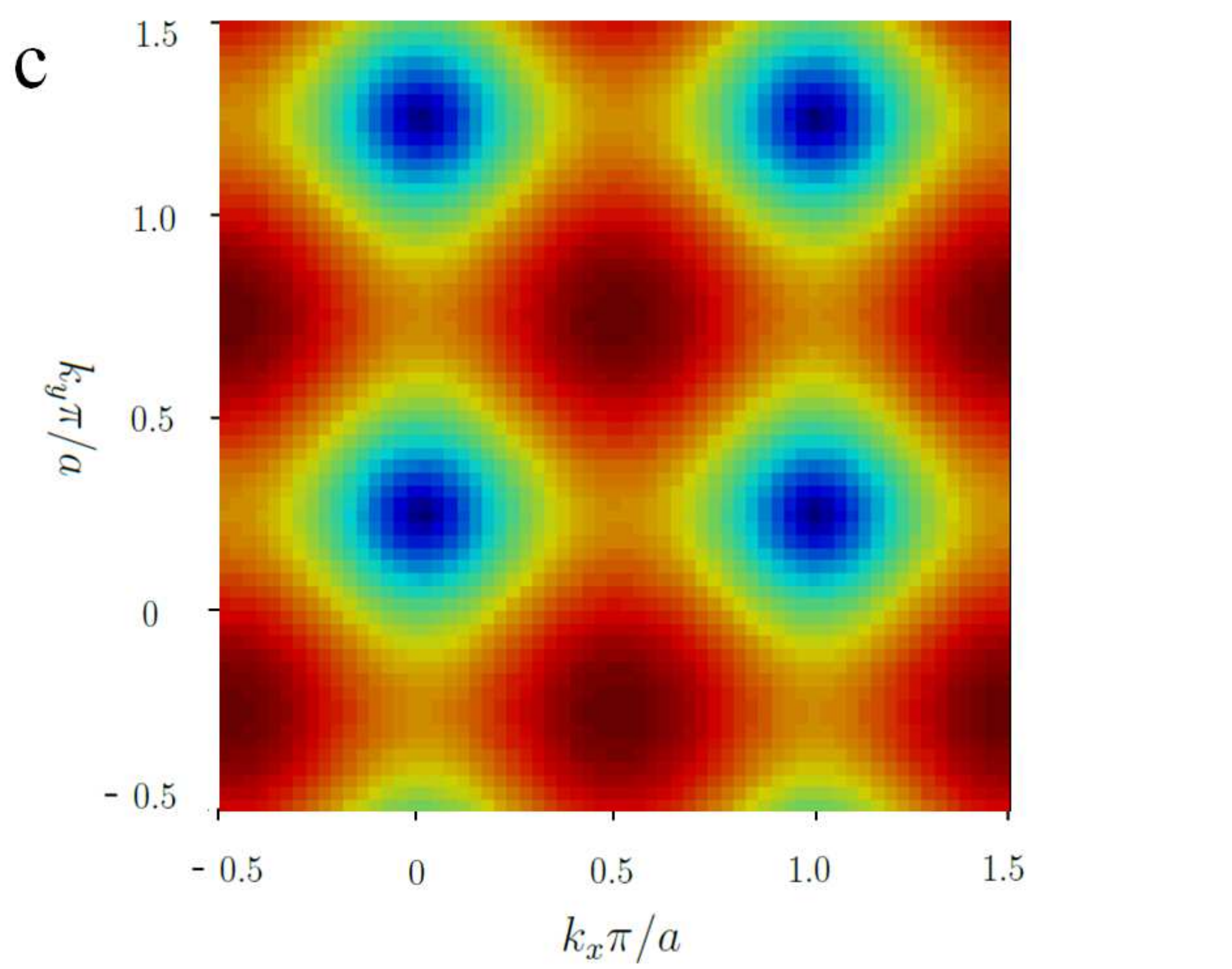} 
\label{fig:diracpoints-c} } 
\subfloat{\includegraphics[width=0.5\columnwidth,height=0.36
\columnwidth]{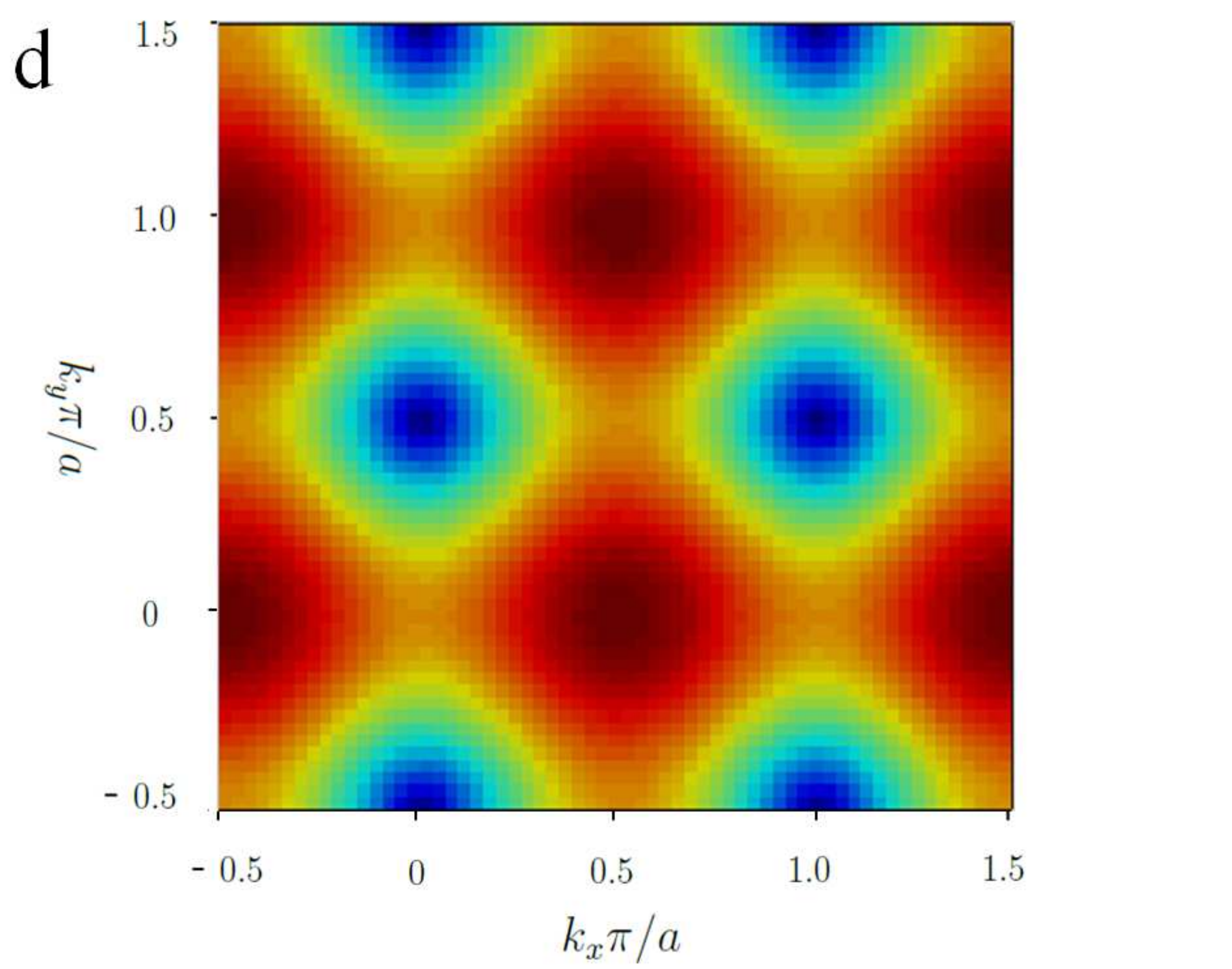} 
\label{fig:diracpoints-d} }
\caption{A three dimensional view of the energy spectrum plotted for a purely ($\Phi = 0$ non-Abelian gauge field. The strength of SOC is (a) $\alpha = \pi/2 = \beta$; (b) $\alpha = \pi/2 + 0.25, \beta = \pi/2 - 0.25$. The 
surface plot is an intensity map of the energy difference between $E_+$ and $E_-$. The four green spots on the surface correspond to the four (Bosonic) Dirac points (at the zone centers) where the energy gap between 
the two bands vanishes. The red band and the blue band correspond to $E_+$ and 
$E_-$, respectively. $W$ is the maximum band-gap, that occurs at the zone boundaries. In (c)-(d) the location of the Dirac points on the momentum space are shown for $2m\pi \Phi=$ 0.75 and 1.5, respectively. With increasing $\Phi$ the Dirac points move along +ve $k_y$ axis. }
\label{fig:diracpoints} 
\end{figure}

It must be emphasized that such massless bosonic quasiparticles which mimic the massless dirac fermions 
in relevant fermionic systems \cite{DiracExpt} arise in this system  as a consequence of the spin-1/2 nature of the bosons. Such spin-1/2 bosons have no natural analogue because of Pauli's spin-statistics theorem. However, this constraint can be lifted by synthetic symmetries \cite{Leggett} and synthetic bosonic (pseudo) spin-half system can be realized \cite{reviewSOC}. After the preliminary proposals on simulation of Dirac fermions in cold atom system \cite{Duan-Dirac} they were soon realized experimentally \cite{DiracExpt}, using density profile measurement methods or Bragg spectroscopy. Similar techniques may also be exploited to observe the bosonic quasiparticles that follows massless Dirac equation. 

As evident from eq. \eqref{diracspec}, the effect of an Abelian field would be to move these points on the momentum space (see Fig. \ref{fig:diracpoints-c}, \ref{fig:diracpoints-d}). With finite Abelian field there also emerges a Hofstadter spectrum as discussed previously. This can be verified by plotting the energy as a function 
of the  abelian ( magnetic) flux \cite{Kubasiak}.  For the same system here in   Fig. (\ref{fig:diracpoints-c}) and (\ref{fig:diracpoints-d}) we plotted the energy  against the Bloch momentum for a given value 
of the Abelian flux to show the location of the Dirac points. From the eq. \eqref{diracspec} it is also suggestive that with the use of a spatially modulated Abelian flux one may control the separation between the Dirac points. Motion and merging of Dirac points has also been very interesting as they lead to topological phase transitions \cite{Dirac-Phases}. One can also switch on the interaction and study its effects on the spectrum \cite{Interaction-Dirac}.

\subsection{Emerging Magnetic Orders}
\label{sec:Magnetic Phases}

In this subsection we discuss about the various magnetic orders that arise in the ground state of the Hamiltonian in eq. \eqref{eq:TB}. This can be done by mapping this Hamiltonian to an effective spin Hamiltonian - one treats the interaction part of eq. \eqref{eq:TB} as the zeroth-order Hamiltonian and then the hopping part ($\mj \hB$) is treated perturbatively to get the effective spin Hamiltonian matrix elements. We do not discuss he full method here, this can be found in \cite{Altman, Kuklov, Lukin, Auerbach}. Using such analysis the effective spin Hamiltonian of a spin-orbit coupled BEC in a classical optical lattice was already obtained in \cite{Trivedi, Sengupta, Radic, Cai}. We realize that the mathematical structure of our effective eBHM Hamiltonian in eq. \eqref{eq:TB} is same to that considered in \cite{Trivedi, Sengupta, Radic, Cai}, provided we switch off the Abelian field part. Since we have considered a cavity induced quantum optical lattice, instead of the hopping amplitude  $t$  in a classical optical lattice, which was the case studied in those works, here we have a rescaled hopping parameter $\mj$, which essentially captures the information of the quantum light. Thus in the parent Hamiltonian of  refs. \cite{Trivedi, Sengupta, Radic, Cai}, if we substitute $\mj$ in place of $t$ we arrive at the same conclusion. In fact, since $\mj$ can be controlled by means of the cavity parameters thus one can also maneuver the entire phase diagram by suitably adjusting these parameters. 

Thus we consider the spin-Hamiltonian obtained in \cite{Radic} and directly substitute $\mj$ in place of $t$ to obtain :
\bea
\hh_{spin} &=& \hh_H + \hh_A + \hh_D , \nn \\
\hh_H &=& -\sum_i \mh \vec{S}_i \cdot (\vec{S}_{i + \delta_x} + \vec{S}_{i + 
\delta_y}) , \nn \\
\hh_A &=& -\sum_i \ma ( S^x_i S^x_{i + \delta_x} + S^y_i S^y_{i + \delta_y}) , \nn \\
\hh_D &=& -\sum_i \md (\vec{S}_i \times \vec{S}_{i + \delta_x} \cdot \hat{x} + 
\vec{S}_i \times \vec{S}_{i + \delta_y} \cdot \hat{y}) ,
\eea
Here $\vec{S_i}$ are the isospin operators at site $i$: $\vec{S_i} = \frac{1}{2} \sum_{s,s'} \hb^\dag_{s i} \vec{\sigma}_{s s'} \hb^{\pdag}_{s' i} $. Each component of the isospin operator are $S^x_i = (\hb^\dag_{\upa , i} \hb^{\pdag}_{\dna , i} + \hb^\dag_{\dna , i} \hb^{\pdag}_{\upa , i} )/2$, $S^y_i = (\hb^\dag_{\upa , i} \hb^{\pdag}_{\dna , i} - \hb^\dag_{\dna , i} \hb^{\pdag}_{\upa , i} )/ 2i$, $S^z_i = (\hb^\dag_{\upa , i} \hb^{\pdag}_{\upa , i} - \hb^\dag_{\dna , i} \hb^{\pdag}_{\dna , i} )$. And $\mh = \frac{4\mj^2}{U} \cos(2\alpha)$, $\ma = \frac{8\mj^2}{U} \sin^2 
\alpha$, and $\md = \frac{4\mj^2}{U} \sin(2\alpha)$ are the spin interaction 
strengths. The effective spin Hamiltonian $\hh_{spin}$ is a combination of 
two-dimensional Heisenberg exchange interactions ($\hh_H$), anisotropy 
interactions ($\hh_A$), and Dzyaloshinskii-Moriya interactions ($\hh_D$) \cite{DMInteraction}. These terms collectively stabilize the following orders \cite{Radic}: ising ferromagnets (zFM), antiferromagnets (zAFM), Stripe phase, Spiral phase (commensurate with 3-sites or 4-sites periodicity, respectively denoted as 3-Spiral and 4-Spiral), and the vortex phase (VX).

A detailed discussion of these phases can be found in \cite{Radic}. We discuss these phases briefly. A schematic of the spin configurations of these phases are given in the insets of Fig. \ref{fig:CavSpecRII}. The zFM order is a uniformly ordered phase where all the spins are aligned along the z-axis, however in the zAFM phase the direction of the spin vectors alternate as parallel or anti-parallel to the z-axis. There is a subtle difference between the stripe phase and the zAFM: in the stripe phase, along a given axis on the xy-plane all spins are up but for the other axis they alternate as up and down. In zAFM they the spins alternate along both the axes. Two types of spiral waves appear for this system. In both the cases, all the spins along one axis on xy-plane are parallel, however along the other axis, the spin vectors make an angle with the z-axis which changes (starting from 0) as we move along the axis. However, there exists a period in number of lattice sites after which the angles are repeated like wave. In 4-spiral, 4 sites make one period: the angles progress with site as $\pi, \pi/2, 0, -\pi/2, \pi ...$. In 3-spiral, 3 sites make one period: the angles progress with site as $\pi, \pi/3, -\pi/3, \pi ...$. The vortex phase is one of the XY phases, in which all the spin vectors lie on the XY plane. In section \ref{sec:CavitySpectrum} we will see how we can detect all these phases. 

\section{The Cavity Spectrum for the Magnetic Phases}
\label{sec:CavitySpectrum}

In the preceding section, we discussed the spectrum of the non-interacting SOC bosons in a cavity induced quantum optical lattice potential. Now we switch on the atom-atom interaction. As pointed out in sec. 
\ref{sec:Magnetic Phases}
this causes appearance of various magnetic orders in the many body quantum mechanical ground state. These orders have been studied in cold atomic systems, in presence \cite{Trivedi, Sengupta, Radic, Cai} or absence \cite{Altman, Kuklov, Lukin} of SOC. The many body wavefunction has an orbital part and a spinorial part and the magnetic orders are characterized by the spinorial part of the wavefunction. Detection of various phases in the orbital part of the wavefunction, through the cavity spectrum was carried out in \cite{Mekhov}. In our work we propose a method which enables us to probe the spinorial part of the wavefunction (hence the magnetic orders) with the help of the cavity spectrum. 

We define the cavity spectrum is the steady state outcoming (leakage) photon number which is obtained from \eqref{eq:Photon} by setting $\partial_t \hat{a} = 0$ as: 
\beq
n_{ph} = \expect{ \ha^{\dag (s)}\ha^{(s)}}_\Psi = \frac{\eta^2}{\kappa^2 + 
(\Delta_c' - U_0 J_1 \expect{\hB}_\Psi)^2} .
\label{eq:nph}
\eeq
This equation is non-linear \cite{Larson} in terms of photon density $n_{ph}$ since the tunneling parameters, $J_0$ and $J_1$ are dependent upon the depth of the optical lattice potential, $V_0 = U_0 n_{ph}$. Essentially cavity induces a feedback mechanism (of cavity light) causing the cavity spectrum to nonlinearly depend on $n_{ph}$ through this modified Lorentzian \cite{Meystre-Book}. In addition, the spectrum is also dependent upon the state $\ket{\Psi}$ through the expectation value of the hopping operator $\expect{\hB}_\Psi$. This dependence is pronounced only when $J_1$ is finite. In further discussions we will show how this dependence can be used to probe the spinorial part of the quantum many-body ground state wavefunction.

The ground state of the BH model is controlled by the value of $t/U$ \cite{Fisher, Trivedi}. As the depth of the potential well increases, the ground state changes from a super fluid (SF) to Mott insulator (MI) state. To simplify our discussion we assume that the orbital (optical lattice site) part of the  wavefunction corresponds to a Mott insulator state with one atom per lattice site. In absence of any (synthetic) gauge field, for a 2D lattice, this phase boundary occurs at $U = 4(3+2\sqrt{2})t$, which can be obtained from mean-field calculations \cite{DMFT}. The presence of (synthetic) Abelian gauge field further localizes the atoms and the phase boundary gets shifted towards a larger value of $t/U$ or, a more shallow lattice \cite{Grass}. So we confine our discussion to lattice depth larger than $20E_r$.

We further divide the MI regime into two regions separated at a potential depth of $25E_r$ (see Figure \ref{fig:CavSpec}a). In one region of the depth values the $J_1$ vanishes, hence it becomes impossible to probe the spinorial part of ground state through the cavity spectrum. In the other region the $J_1$ is finite, enabling us to probe the ground state. We name these regions as region I: Shallow MI regime ($\lesssim 25 E_r$), where $J_1 \neq 0$ and hence the equation \eqref{eq:nph} is highly non-linear; region II: Deep MI regime ($\gtrsim 25 E_r$), where $J_1$ is approximated to $0$ and the non-linearity in $n_{ph}$ enters only through $J_0$.
\begin{figure}[]
\centering
\subfloat[]{\includegraphics[width=0.9\columnwidth, height=0.75 \columnwidth]{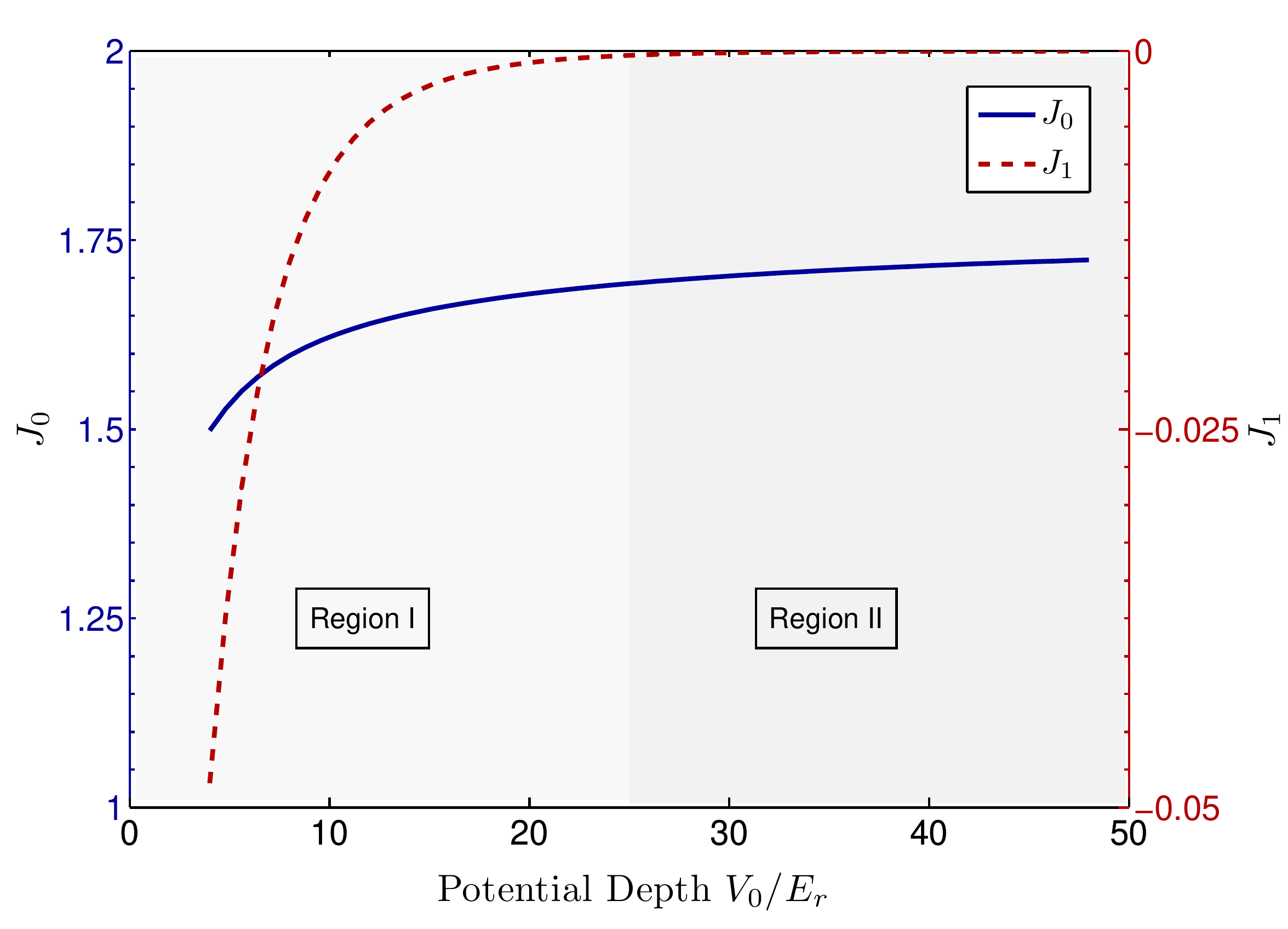}
\label{fig:CavSpec-a}} \\
\subfloat[]{\includegraphics[width=0.5\columnwidth, height=0.45\columnwidth]{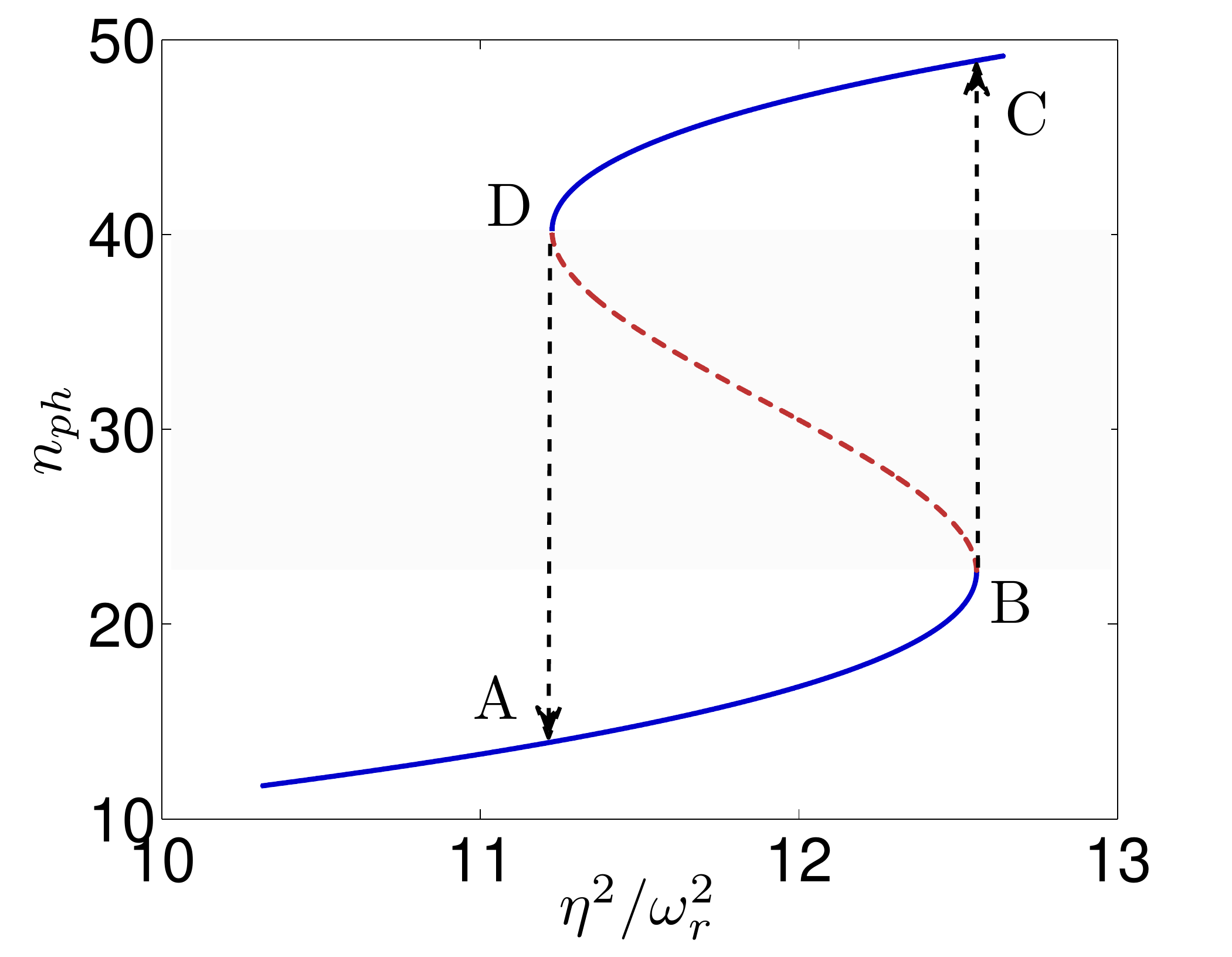}
\label{fig:CavSpec-b}}
\subfloat[]{\includegraphics[width=0.5\columnwidth, height=0.45\columnwidth]{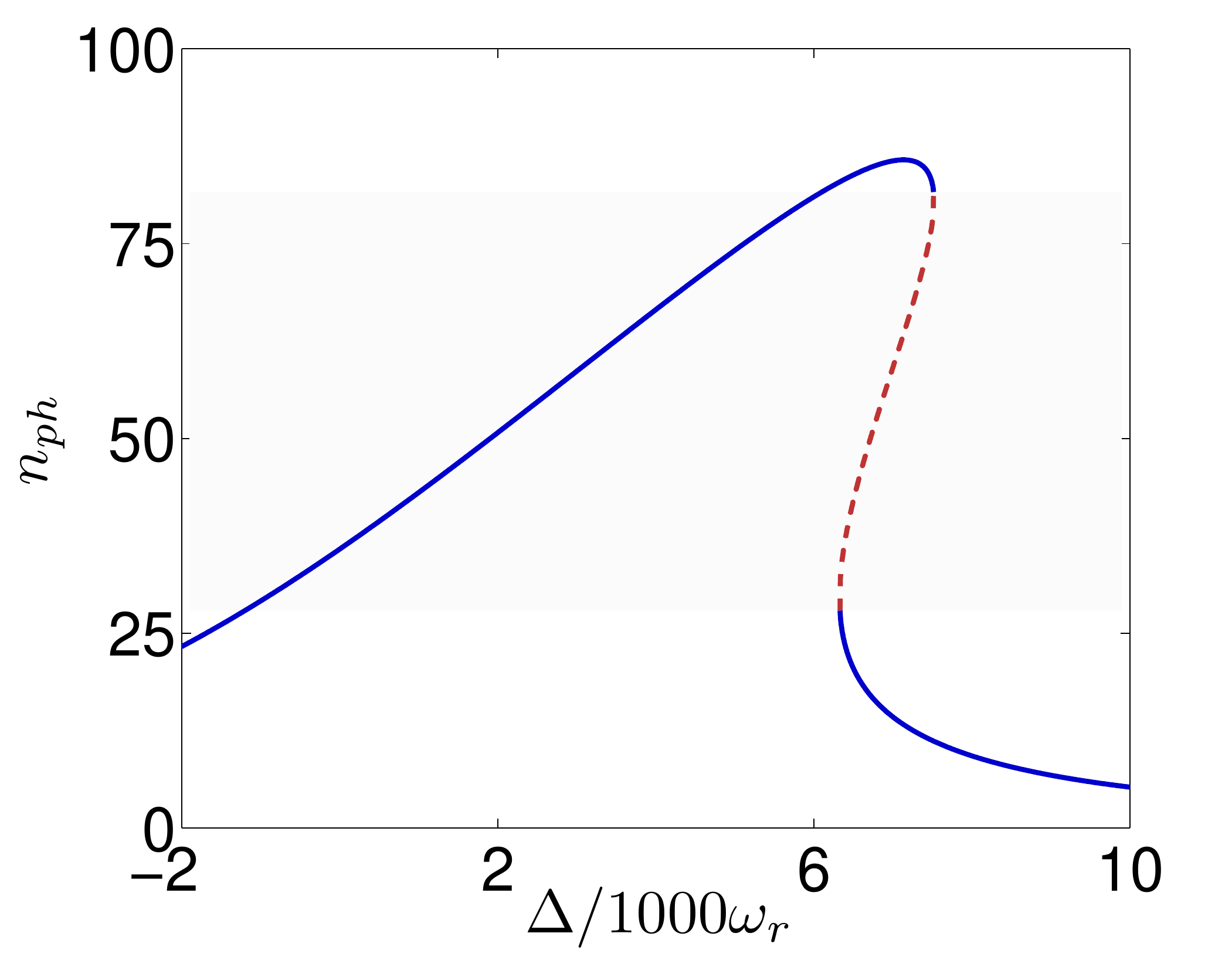}
\label{fig:CavSpec-c}}
\caption{{\bf a}. The variation of overlap integral elements with potential 
depth. We study the variation in two regions, separated at $V = 25 E_r$; Cavity Spectrum for a deep lattice (Region I), for a $6 \times 6$ lattice, $\{U_0, \kappa \}= \{ 12, 1 \} \omega_r $, {\bf b}. with pump amplitude $\eta$ for $\Delta_c = 5000 \omega_r$; {\bf c}. with detuning $\Delta_c$ for $\eta = 6 \omega_r$. The red dotted lines are the unstable regions of photon count.}
\label{fig:CavSpec}
\end{figure}

Lets first consider region II. As evident from Figure \ref{fig:CavSpec-a} in this 
region $J_0$ vs $V_0$ can be approximated by a linear function ($J_0 = aV_0+b$) 
and $J_1$ can be assumed to be zero. The variation of $n_{ph}$ with respect to pump amplitude $\eta^2$ is shown in Figure \ref{fig:CavSpec-b} and that with respect to detuning $\Delta_c'$ is shown in Figure \ref{fig:CavSpec-c}. There exists a bistable region in the spectrum which is shown by red dashed line. In the strong MI regime the atoms get tightly localized at their site resulting in a negligible hopping amplitude. The atoms can sense the presence of the abelian or non-abelian field only through the hopping term, and now since the hopping amplitude is almost negligible the cavity spectrum is insensitive to the abelian or non-abelian gauge field. 

As the pumping amplitude $\eta$ decreases the photon number decreases (see Figure \ref{fig:CavSpec-b}, however at a certain point (point D) the photon number abruptly drops to a very small value (point A), hence the lattice suddenly becomes very shallow. This causes a phase transition from Mott insulator to superfluid phase. Similarly, as $\eta$ increases the photon number also increases, so does the lattice depth as well. At the point B it suddenly jumps to a large value of $n_{ph}$ (point C) hence a phase transition from super fluid to mott insulator occurs. This is an instance of bistability driven driven phase transition, which was previously pointed out in \cite{Larson}, \cite{Meystre} in different contexts. Points B or D are often referred to as turning points or critical points. When the photon number gets lowered one might end up at a super fluid phase or one might stay in the shallow MI region. So to determine the phase exactly one needs to obtain the exact phase diagram and locate the appropriate turning points. We do not extend this discussion further.

Now we turn to the case of shallow MI regime (or region I). We separate the following section where we show that in this region it is feasible to probe the ground state of the SOC BEC through the cavity spectrum. When $J_1 \neq 0$, the Lorentzian in \eqref{eq:nph} can sense the presence of the magnetic orders through $\expect{\hB}$. In section \ref{sec:Magnetic Phases} we have already introduced and discussed briefly the magnetic orders that prevails in such a system. 

Before getting to our results, it is worthwhile to point out that after the realization of spin-orbit coupling for bosonic clouds 
\cite{reviewSOC} or condensate \cite{Spielman1} by Spielman's group the phase 
diagram of such a system was theoretically obtained by various groups in 
\cite{Trivedi,Sengupta,Radic,Cai}. Experimental verification of these phases 
might not be very trivial, most importantly detecting all the emergent phases 
using a single experimental setup is a formidable  task. So far, the method 
of spin structure factor measurement through Bragg spectroscopy \cite{Bragg} has 
been commonly used. Other methods include measurement of spatial noise 
correlations \cite{AltmanNoise}, polarization-dependent phase-contrast imaging 
\cite{Kurn-Magnetization}, direct imaging of individual lattice sites 
\cite{SingleSite} etc.. However, each of these techniques come with 
their own set of  complications. 

Extending the idea which 
was originally espoused for BEC without spin degrees of freedom
\cite{Mekhov} here we propose a differentl scheme of experiment where
such magnetic orders can be ascertained without making a direct measurement 
on the atomic system. The relation between such approach and 
 "quantum nondemolition measurement" technique  was also discussed \cite{SelfOrganization2, Ritsch1, Supersolid}. The method facilitates the detection all possible phases arising in 
the Mott regime of a SOC BEC and this can also be extended to the superfluid 
(SF) regime.

To this purpose we work out the values of $\expect{\hB}$ and obtain the cavity spectrum. Following \cite{Girvin} the wave function for various orders can (in the Mott 
phase only) be written as
\beq
\ket{\Psi_{MI}} = \prod_{i \in A , j \in B} \ket{\psi_A}_i \ket{\psi_B}_j ,
\label{eq:MI}
\eeq
with site indexes $i, j$ and $\ket{\psi_{A,B}} = \cos \frac{\theta_{A,B}}{2} 
\ket{\upa} + e^{ i \phi_{A,B}} \sin \frac{\theta_{A,B}}{2} \ket{\dna} $. The 
entire lattice is divided into two sub-lattices $A, B$ and we assume alternating 
sites belong to different sub-lattices. The parameters $\theta , \phi$ are 
projection angles in the internal spin space. We assume there are exactly equal 
number of lattice sites in sub-lattices $A$ and $B$, hence the total number of 
sites is $K^2$ even, also assuming unit filling we set $K^2 = N_0$. Please note 
$K$ was earlier used to denote the wave number of the cavity photon and here we use the 
same notation for a different thing. In the 
appendix \ref{app:tunnel} we calculate the expectation value of the tunneling 
operator, $\expect{\hB}$ for various magnetic orders and summarize in the Table 
\ref{tbl}. This will be the basis of further discussions.
\begin{table}[ht]
\caption{\label{tbl} Expectation of the hopping operator and the steady-state 
photon number for different phases in the MI state.}
\begin{ruledtabular}
\begin{tabular}{ll}
{\bf Order}  & \quad $\expect{\hB}$ \\ \hline 
zAFM & $0$ \\
Stripe & $2K(K-1)\cos \beta$ \\ 
VX & $K(K-1)(\cos \alpha + \cos \beta)$ \\ 
3-Spiral & $3K(K-1)(\cos \alpha + 4\cos \beta)/8$ \\ 
4-Spiral & $K(K-1)(\cos \alpha + 3\cos \beta)/2$ \\ 
zFM & $2K(K-1)(\cos \alpha + \cos \beta)$ \\
\end{tabular}
\end{ruledtabular}
\end{table}
\begin{figure*}
\centering
\subfloat{ \includegraphics[width=0.45\textwidth, height= 0.35\textwidth]{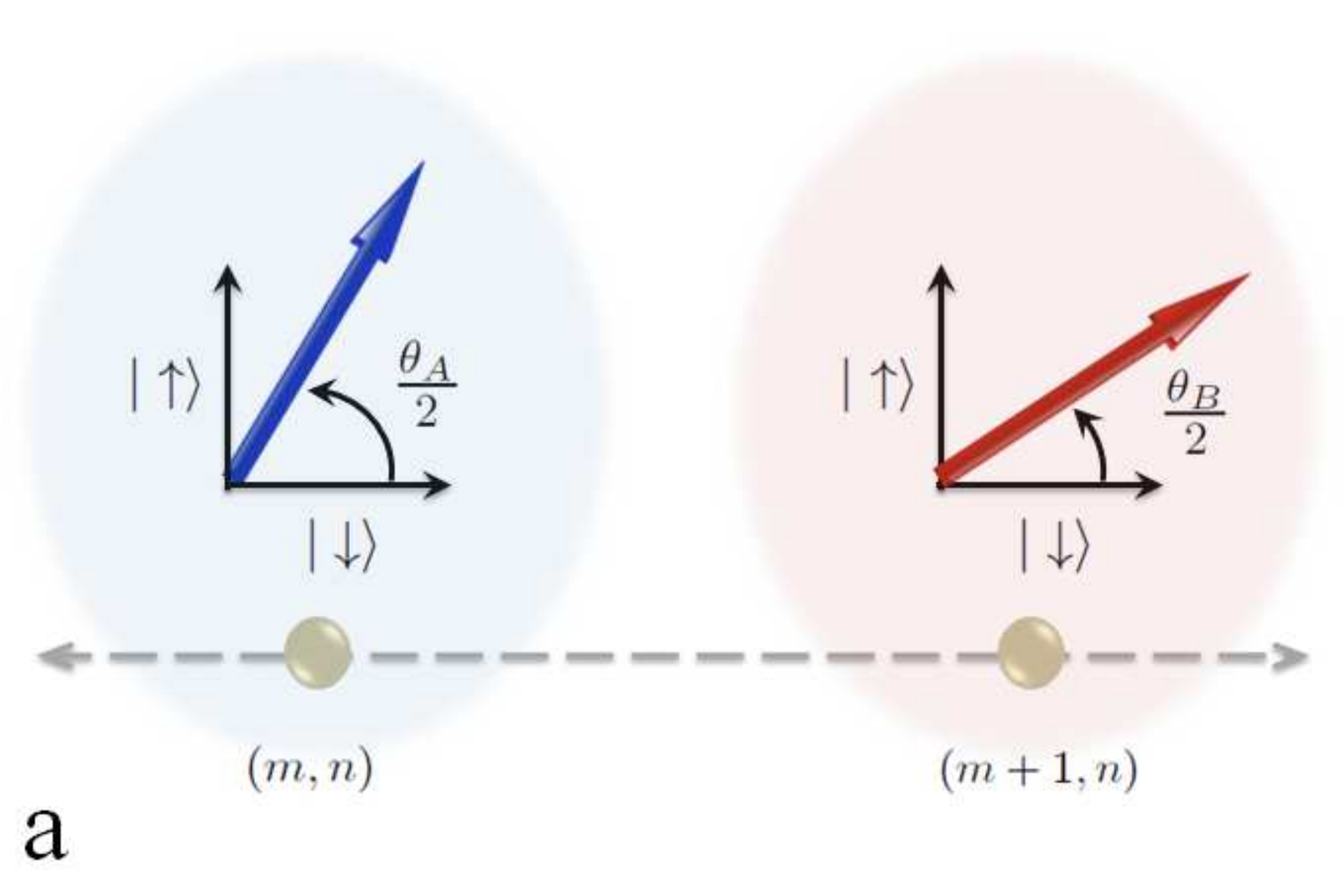} 
\label{fig:CavSpecRII-a}}
\subfloat{\includegraphics[width=0.45\textwidth, 
height=0.4\textwidth]{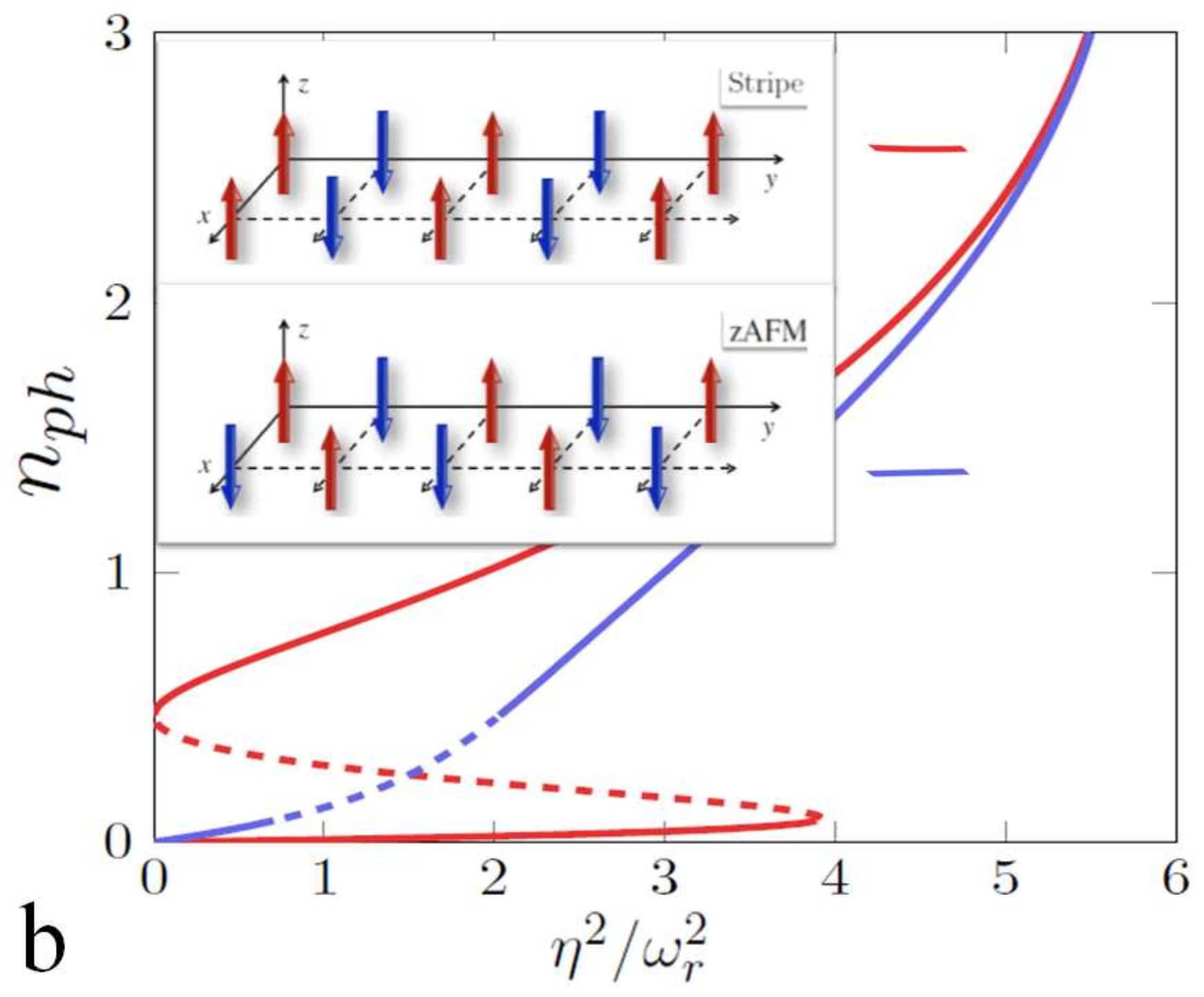} 
\label{fig:CavSpecRII-b}} \\
\subfloat{ \includegraphics[width=0.45\textwidth, 
height=0.4\textwidth]{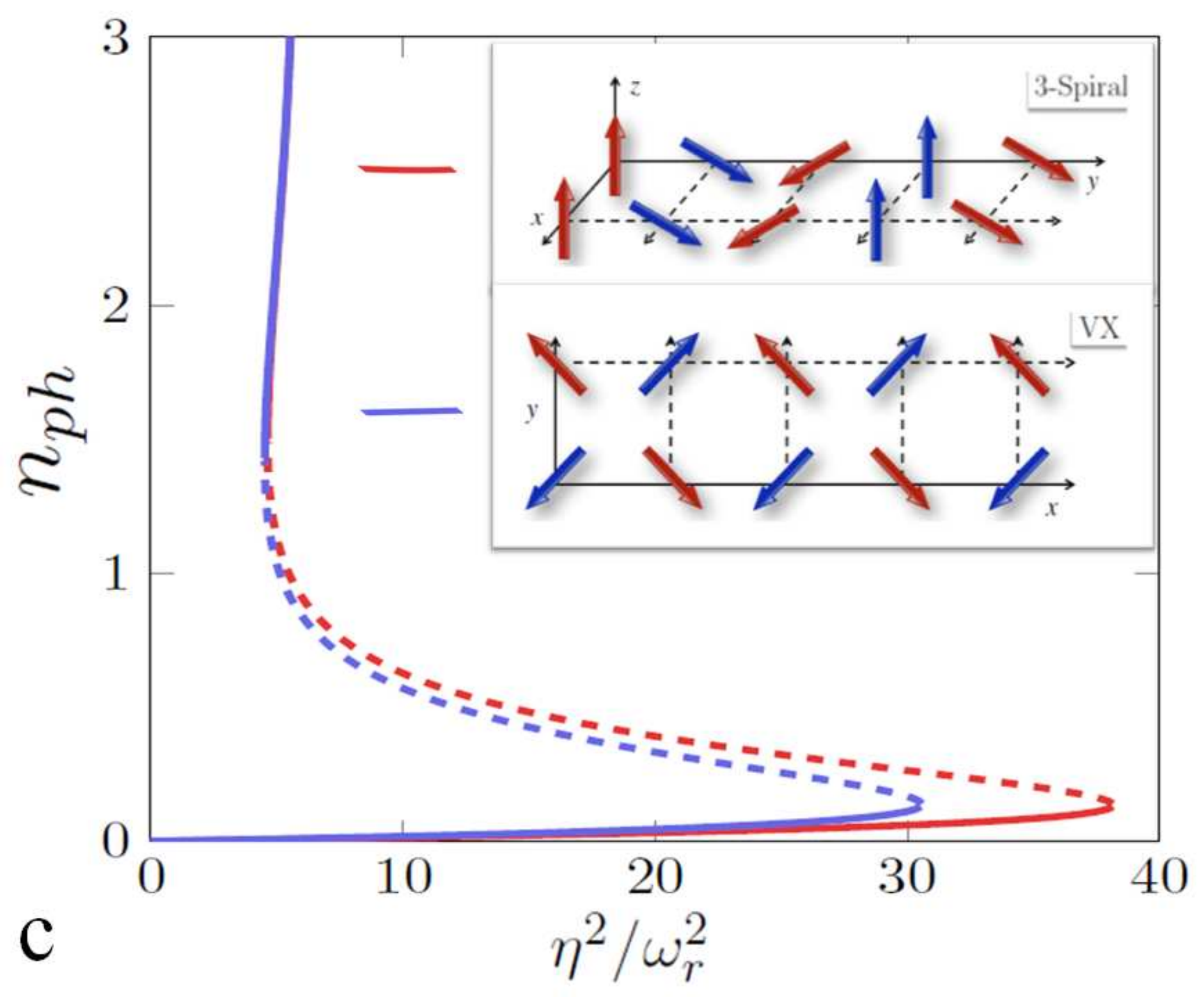} 
\label{fig:CavSpecRII-c}}  \qquad
\subfloat{ \includegraphics[width=0.45\textwidth, 
height=0.4\textwidth]{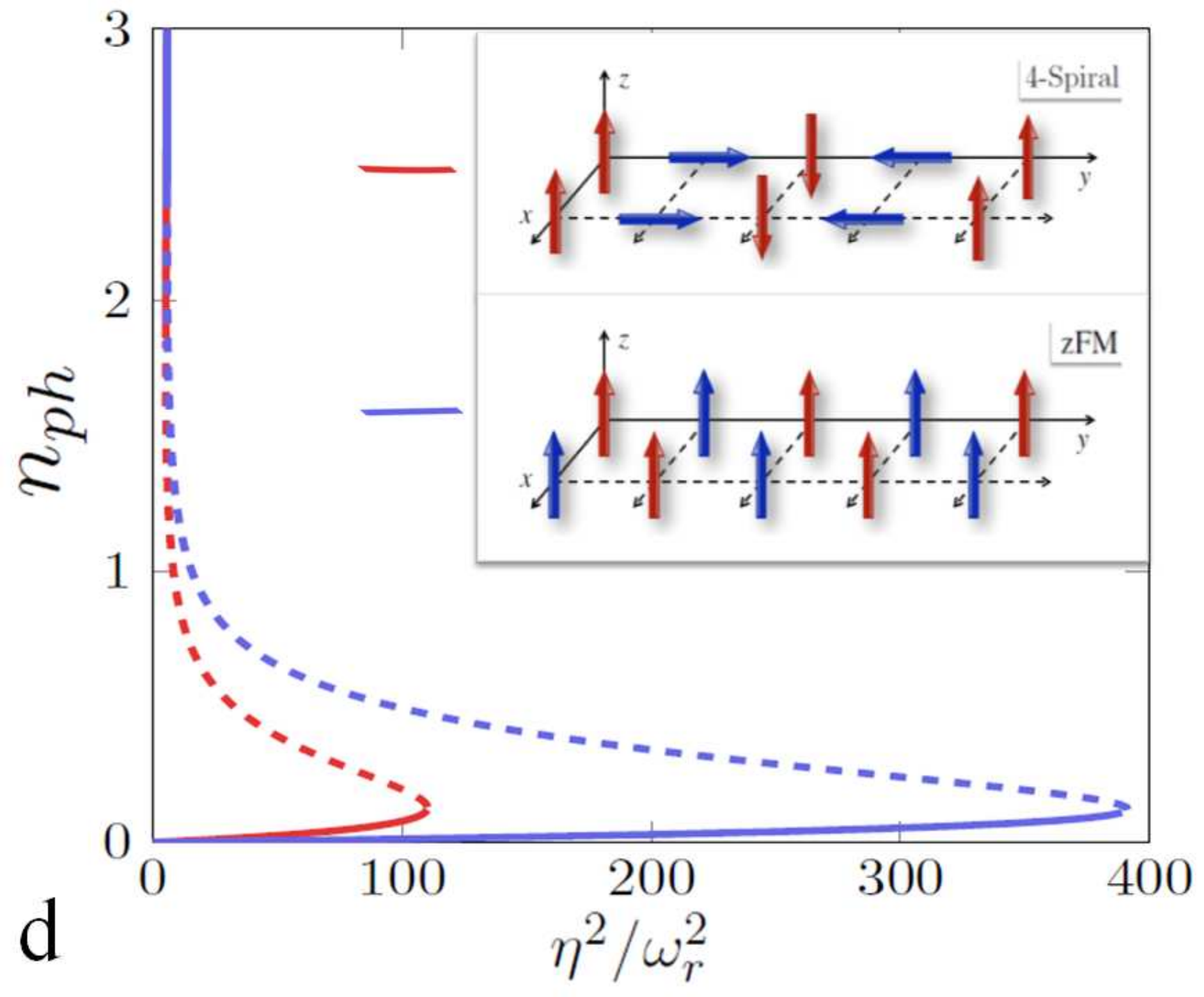} 
\label{fig:CavSpecRII-d}} 
\caption{(a) The spin vectors in internal spin spaces of two neighboring sites; (b)-(d)Cavity spectrum for different phases in the MI region for different 
non-Abelian flux insertions. The SOC strength for all the phases are $(\alpha, 
\beta)/\pi$ = (0.01,0.01) zFM; (0.2, 0.2) 4-Spiral; (0.3,0.3) 3-Spiral; 
(0.5,0.5) Stripe; (0.34, 0.34) VX. Note the turning points are highly dependent 
upon the phases. The dotted part shows the unstable region of the spectrum. The red and blue legends correspond to the magnetic order, shown in boxes.}
\label{fig:CavSpecRII}
\end{figure*}
We can distinguish between different magnetic orders because each order 
can now be associated with a corresponding $\expect{\hB}$, hence a 
caviy spectrum, {\it provided} there is non-vanishing z-axis component
of the spin vector (the reason will be clear later on). Thus one can not distinguish between any of the XY phases, such as the vortex phase or the anti-vortex phase etc. However, the other various magnetic orders, which can arise in a spin-orbit coupled system through experimental control of the free parameters ($\alpha, \beta $) \cite{Radic} or ($\alpha, \lambda$) \cite{Trivedi, Cai} can be well distinguished. 

The cavity spectra for each of these orders are obtained in Figure \ref{fig:CavSpecRII}. The spin-orbit coupling strength $(\alpha , \beta)$ ch                                                                                                                                                                                                                                                                                                                         osen for a particular order is such that, that specific order gets stabilized \cite{Radic}. As we gradually increase the pump value the photon number gets increased, but at the turning point ($\eta_c$) it suddenly jumps to a higher value of photon number, since the photon intermediate count corresponds to the unstable region. Clearly, the behavior of the spectra for different orders are different, specifically the value of $\eta_c$ varies widely. The zAFM will not show any such jump, and the stripe phase will have a very small value of $\eta_c$. For zFM phase $\eta_c$ will always be the largest and for 4-spiral phase it would be quite comparable with the $\eta_c$ of zFM. The XY phase and 3-spiral have there $\eta_c$ always in between these two extremes. 

The above discussion is supported by the following observation. In Figure \ref{fig:CavSpecRII-a} the internal spin (by 'spin' we actually refer to 'pseudo-spin') spaces of two neighboring sites are shown as red or blue blobs. The basis vectors of the spin spaces are the eignvectors of $\hat{S}_z$. If a spin vector makes an angle $\theta$ with the z-axis in the real space, then in the spin space it makes an angle $\theta/2$ with the $\dna$ axis. A particular magnetic order is nothing but a specific spatial distribution of these $\theta$ and $\phi$ values. The value of $\expect{\hB}$ is a measure of the probability of spin-dependent hopping across neighboring sites, which hence captures this variation of $\theta$ values over the configuration space. We proceed in the following way (see appendix \ref{app:tunnel} for rigorous derivation): if a spin vector creates an angle $\theta_A$ with the z-axis and the spin vector at the site nearest to it makes an angle $\theta_B$ then in their internal spin spaces they make an angle $\theta_A/2$ and $\theta_B/2$ with $\dna$. Hence the projection of the spin vectors on the $\dna$ axis are $\cos \theta_{A,B}/2$ and that on the $\upa$ axis are $\sin \theta_{A,B}/2$. The probability for a hopping of $\upa$ to $\upa$ (or $\dna$ to $\dna$) is the modulus squared product of the projection lengths along $\upa$ ($\dna$) axes. Hence for hopping of $\upa$ to $\upa$ has a probability of $( \sin \frac{\theta_A}{2}\sin \frac{\theta_B}{2})^2$ and for hopping of $\dna$ to $\dna$ it is $(\cos\frac{\theta_A}{2} \cos\frac{\theta_B}{2})^2$. Since $\upa$ and $\dna$ are orthogonal vectors hopping associated with a spin flip is found to have vanishing $\expect{\hB}$.

To illustrate the implication of the above technique consider the case of zAFM. In zAFM on alternative sites spin vectors are oriented parallel or anti-parallel to the z-axis, i.e. $\theta_A = 0,$ $\theta_B = \pi$. Hence any reordering of the spin vectors (mediated by the cavity light) which do not alter the magnetic order should consist of hopping from $\upa$ to $\dna$ or visa-versa. However, the matrix element $\expect{\hB}$ for such a hopping is zero. Hence $\expect{\hB}_{zAFM} = 0$ (see the Table). Similarly in case of zFM all spin vectors are aligned along the z-axis, i.e. $\theta_A = \pi = \theta_B $. Hence any hopping other than $\upa$ to $\upa$ will have vanishing contribution in $\expect{\hB}_{zFM}$ and $\expect{\hB}_{zFM} \propto (\sin \pi/2 \sin \pi/2)^2$. It must be noted that the value of $\expect{\hB}$ in turn controls the value of $\eta_c$, hence the trend of variation of $\expect{\hB}$ with respect to the phases gets mapped to that in the values of $\eta_c$. The $\cos \alpha$ or $\cos \beta$ are just scaling factors introduced because of SOC. This is the central result of our work. Now we show that other than the phase information the cavity spectrum can also be used to extract the amount of Abelian or non-Abelian flux inserted in the system.

In order to show how the cavity spectra can be used for flux detection we consider the zFM phase, which is stabilized in presence of both an Abelian and a non-Abelian field \cite{Grass}. In presence of an Abelian flux, the expectation value of the tunneling operator for zFM order becomes (see appendix \ref{app:tunnel}) $\expect{\hB}_{FM} = 2 \cos \alpha (K-1) (K + f(K, \Phi))$. The presence of the Abelian flux gives additional phases to the hopping thus resulting in a overall phase factor of $f(K, \Phi) = \frac{\sin(K \pi \Phi)}{\sin(\pi \Phi)} \cos[\pi \Phi (K-1)]$. This function is plotted in Figure \ref{fig:flux-a}. The similarity of the functional form of $f(K, \Phi)$ with that of an N-slit grating function is just because in this case the phases arising due to the presence of this field gets summed over to yield such a function. Evidently the optical lattice acts as a quantum diffraction grating \cite{Adhip, Mekhov}. 
\begin{figure}
\subfloat[]{ \includegraphics[width=0.9\columnwidth, height=0.75\columnwidth]{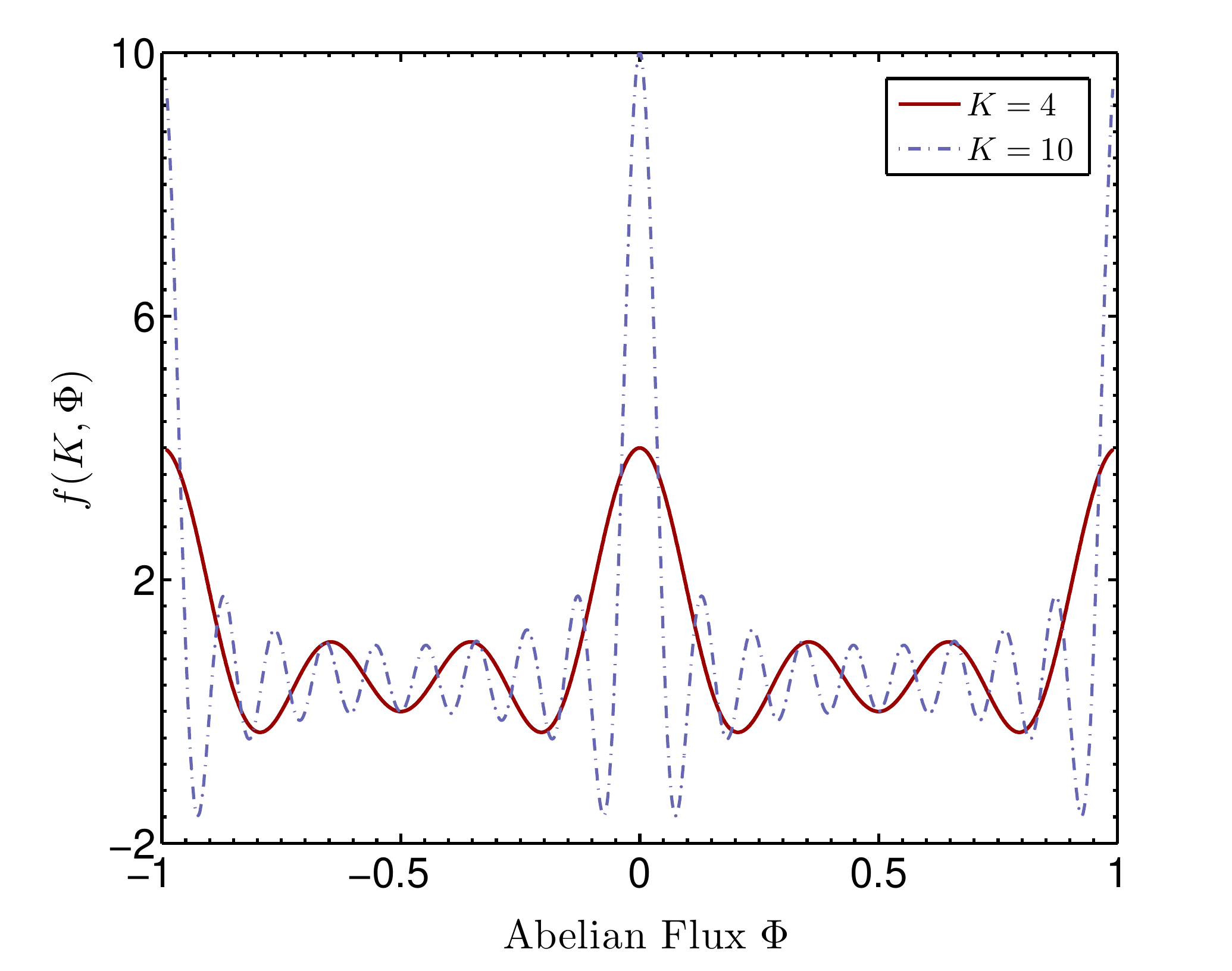} 
\label{fig:flux-a}} \\
\subfloat[]{ \includegraphics[width=0.45\columnwidth, height=0.42\columnwidth]{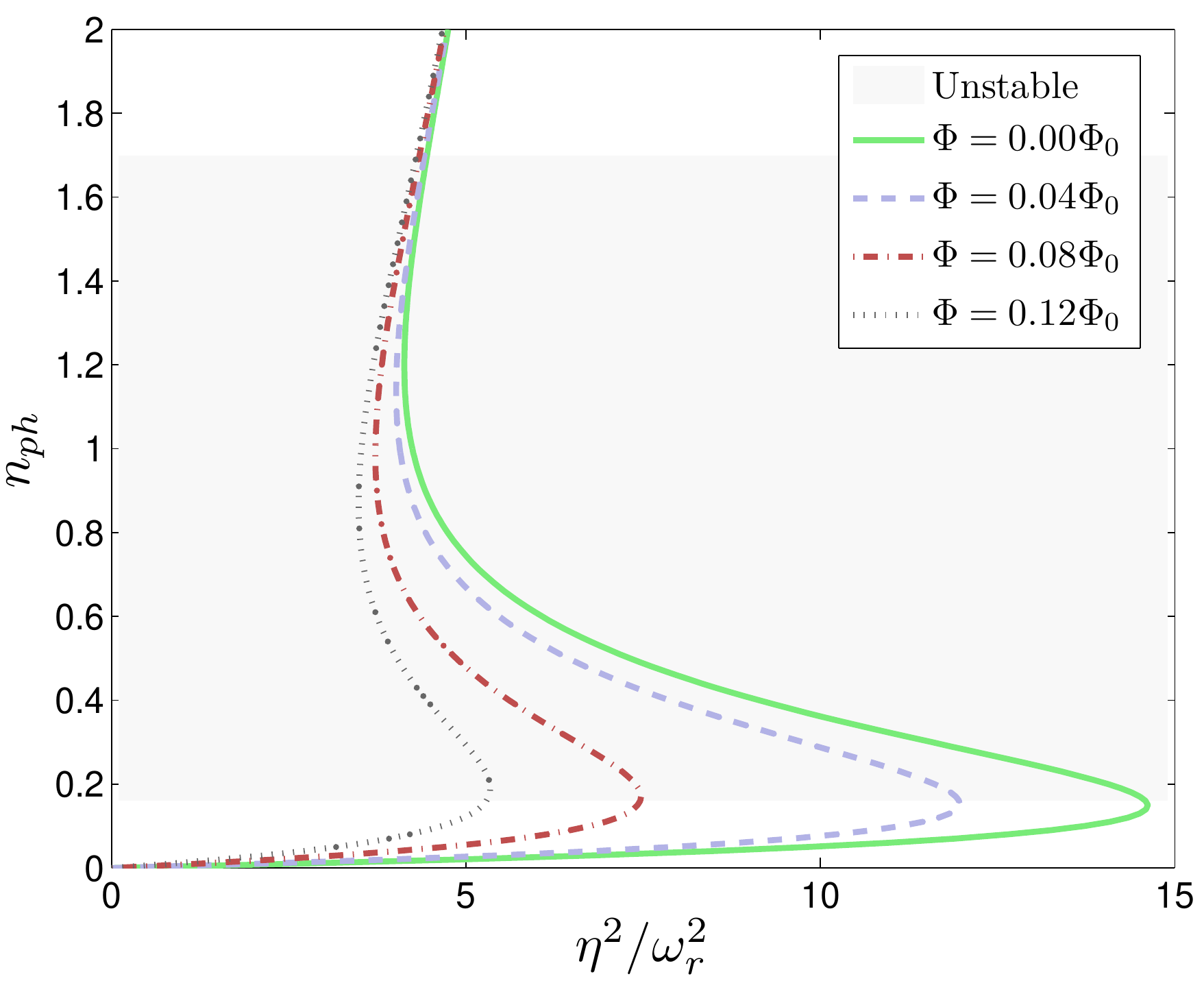}
\label{fig:flux-b} }
\subfloat[]{ \includegraphics[width=0.45\columnwidth, height=0.42\columnwidth]{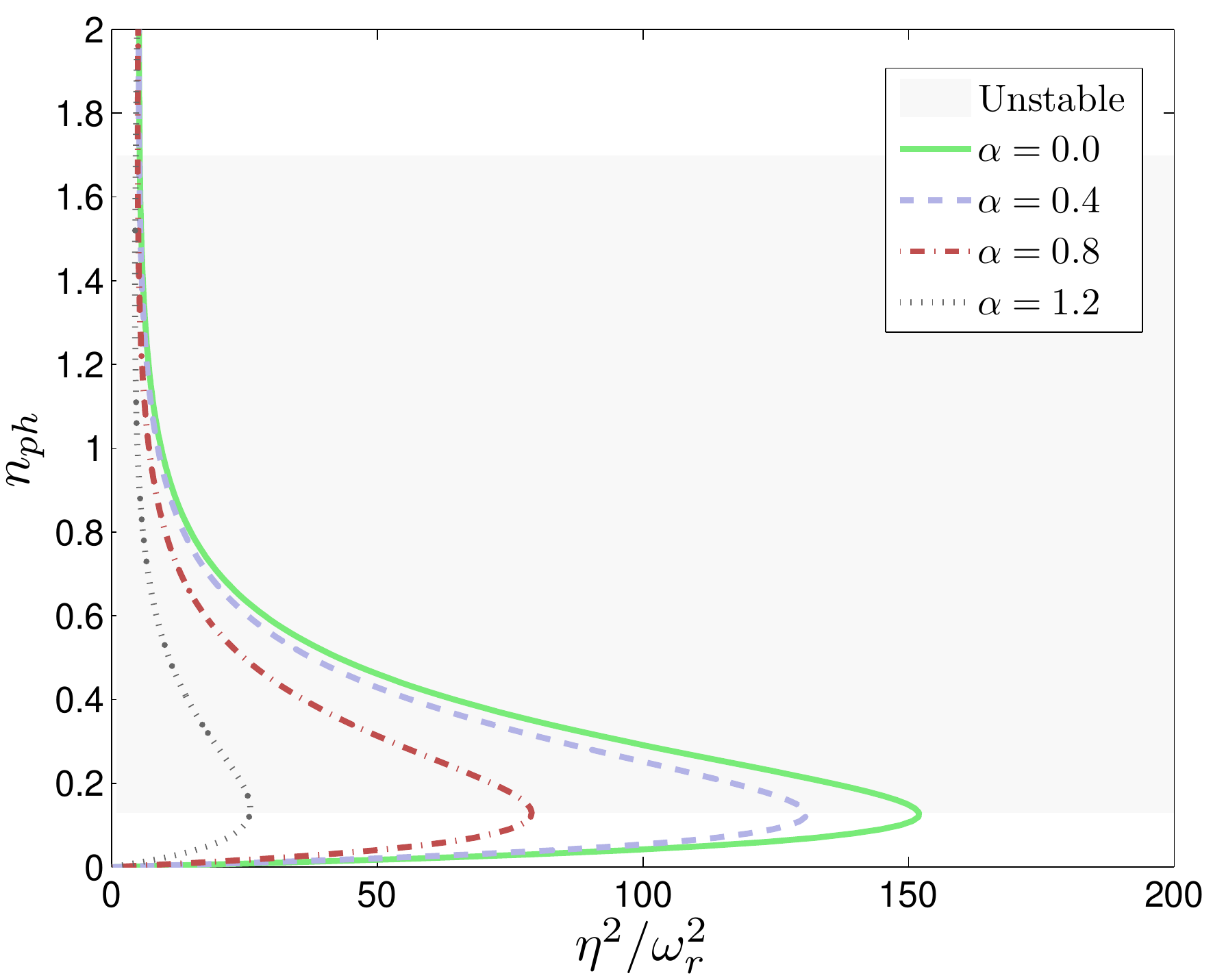} 
\label{fig:flux-c}}
\caption{{\bf a}. The variation of grating function $f(K,\Phi)$ with the 
inserted Abelian flux. The graph legends indicate the size of the lattice. In a 
large lattice limit the grating function does not sense the variation of $\Phi$; 
The cavity spectrum for different - {\bf b.} Abelian fields (with fixed 
non-Abelian field, $\alpha = -\beta = \pi/2 - 0.15$); {\bf c} non-Abelian fields 
(with fixed Abelian field, $\Phi = 0.08 \Phi_0$). The negative slope region is 
the unstable (gray) part of the spectrum.}
\label{fig:flux}
\end{figure}

\section{Conclusion}
To summarize, in this paper we derived an effective moel i.e. eq. \eqref{eq:TB} for SOC-BEC inside a cavity. 
The subsequent analysis based on this effective model indicates a number of very interesting features.
We first studied its spectrum in the non-interacting limit and showed that Dirac-like spectrum arises 
for such ultra cold bosons because of the effective spin-1/2 behavior of this system. 
We also point out that in presence of Abelian flux one can generate highly controllable (through cavity parameters) Hofstadter butterfly spectrum. 

Then we discuss the magnetic phases that arise in the MI type ground state of this Hamiltonian after including atom-atom interaction. Subsequently 
we discuss a technique with which we can probe these magnetic orders through the cavity spectrum. By setting up a lattice, generated by the cavity, we first let the atoms to stabilize in a particular magnetic order. This can be done by adjusting the spin-orbit coupling strengths $\alpha, \beta$ and the inter-atomic interaction strengths $U_{s,s'}$. Then we count the photons leaking out of the cavity as we increase the pump-laser amplitude ($\eta$). We observe at a certain point (the turning point) the photon count suddenly jumps to a very high value. The location of this turning point is characteristic of a specific magnetic order. Hence by locating the turning point we can detect the magnetic phase of the system. Thus our method provides a different way of detecting exotic quantum magnetism in ultra cold condensates.
We would also like to mention that we have only considered the average photon number leaked from the cavity 
as a method to detect the magnetic order inside the cavity. The method can be easily extended by evaluating  quantities like, quadrature measurement, photon number fluctuation, noise spectra and so on \cite{Correlation} and is 
capable of detecting more informations about the quantum phases of SOC-BEC inside the cavity. We hope this 
work will be further extended in this direction and will motivate experiments on Cavity Optomechanics and Cavity 
Quantum Electrodynamics with Spin-Orbit coupled cold gases. 

However, an important issue related to the detection of all these phases is the energy scale of the effective Hamiltonian which gives rise to such phases, i.e. $J^2/U$. Hence the temperature required to realize such orders becomes $\sim J$ which is still not achieved in the current cooling techniques. However, development of new methods of cooling are under progress \cite{Cooling} which is expected to realize such magnetic orders in ultra cold systems. In that context our results provides a very interesting and alternative method of detecting such quantum magnetic 
phases.


\appendix

\section{Frame Transformation}
\label{app:Baker}
We discuss briefly how to arrive from the time-dependent equation \eqref{eq:TimeDep} to a time-independent equation in \eqref{eq:TimeInd}. For this we enter into a rotating frame which induces a unitary transformation $\hat{U}(t) = \exp[i \omega_p t (\hs_{11} + \ha^\dag \ha)]$ and then use Baker-Campbell-Hausdorff lemma to arrive at \eqref{eq:TimeInd}. The lemma reads: 
\beq
e^X Y e^{-X} = Y + [X,Y] + \frac{1}{2!} [X,[X,Y]] + \frac{1}{3!} [X,[X,[X,Y]]] + ... \eeq
For our case $X = i \omega_p t (\hs_{11} + \ha^\dag \ha)$ and $Y = \hh_A + \hh_C + \hh_I$ as obtained in \eqref{eq:TimeDep}. We evaluate the following commutators one by one:
\bea
[X, \hh_A] &=& i \omega_p t [ \hs_{11} + \ha^\dag \ha , \frac{\hat{\bs \Pi}^2}{2m} + \hbar \omega_{12} \hs_{11} + \hbar 
\omega_{13} \hs_{11} ] \nn \\
& = & i \omega_p t  \hbar \omega_{12} [\hs_{11}, \hs_{11}] + i \omega_p t  \hbar \omega_{13} [\hs_{11}, \hs_{11}] \nn \\
&=& 0.
\eea
\bea
[X, \hh_C] &=& i \omega_p t [ \hs_{11} + \ha^\dag \ha , \hbar \omega_c \hat{a}^\dag \hat{a} - i \hbar \eta \big ( \ha 
e^{i \omega_p t} - \ha^\dag e^{-i \omega_p t} \big )] \nn \\
&=& \hbar \eta \omega_p t \Big ( [\ha^\dag \ha , \ha ] e^{i \omega_p t} - [\ha^\dag \ha , \ha^\dag ] e^{- i \omega_p t} \Big ) \nn \\
&=& - \hbar \eta \omega_p t \Big ( \ha e^{i \omega_p t} + \ha^\dag e^{- i \omega_p t} \Big ) 
\eea
\bea
[X, \hh_I] = \hbar g(\bs x) \omega_p t [ \hs_{11} + \ha^\dag \ha ,  \Big ( \hs_{12} \hat{a} - \hs_{21} 
\hat{a}^\dag + \hs_{13} \hat{a} - \hs_{31} \hat{a}^\dag \Big )
\eea
We note the following commutators: $[\hs_{11}, \hs_{12}] = [\ket{1}\bra{1}, \ket{1} \bra{2}] = \ket{1} \bra{2} = \hs_{12} $. Similarly, $[\hs_{11}, \hs_{21}] = - \hs_{21}$, $[\hs_{11}, \hs_{13}] = \hs_{13}$, $[\hs_{11}, \hs_{31}] = - \hs_{31}$. Using these the above equation gets simplified as 
\bea
[X, \hh_I] &=& \hbar g(\bs x) \omega_p t \Big ( \hs_{12} \ha + \hs_{21} 
\hat{a}^\dag + \hs_{13} \hat{a} + \hs_{31} \hat{a}^\dag - \hs_{12} \ha - \hs_{21} \ha^\dag - \hs_{13} \ha - \hs_{31} \ha^\dag \Big ) \nn \\
&=& 0.
\eea
Hence the only non-vanishing commutator is $[X, \hh_C]$. Its higher order commutators can be evaluated similarly, e.g. $ [X, [X, \hh_C]] = i \hbar \eta \omega_p^2 t^2 \Big ( \ha e^{i \omega_p t} - \ha^\dag e^{- i \omega_p t} \Big )$, and so on. Plugging all these commutator values to the Baker's lemma we arrive at equation \eqref{eq:TimeInd}.

\section{The Hopping Operator}
\label{app:Hopping}
\begin{figure}
\centering
\includegraphics[width= 0.85 \columnwidth, height=0.7 \columnwidth]{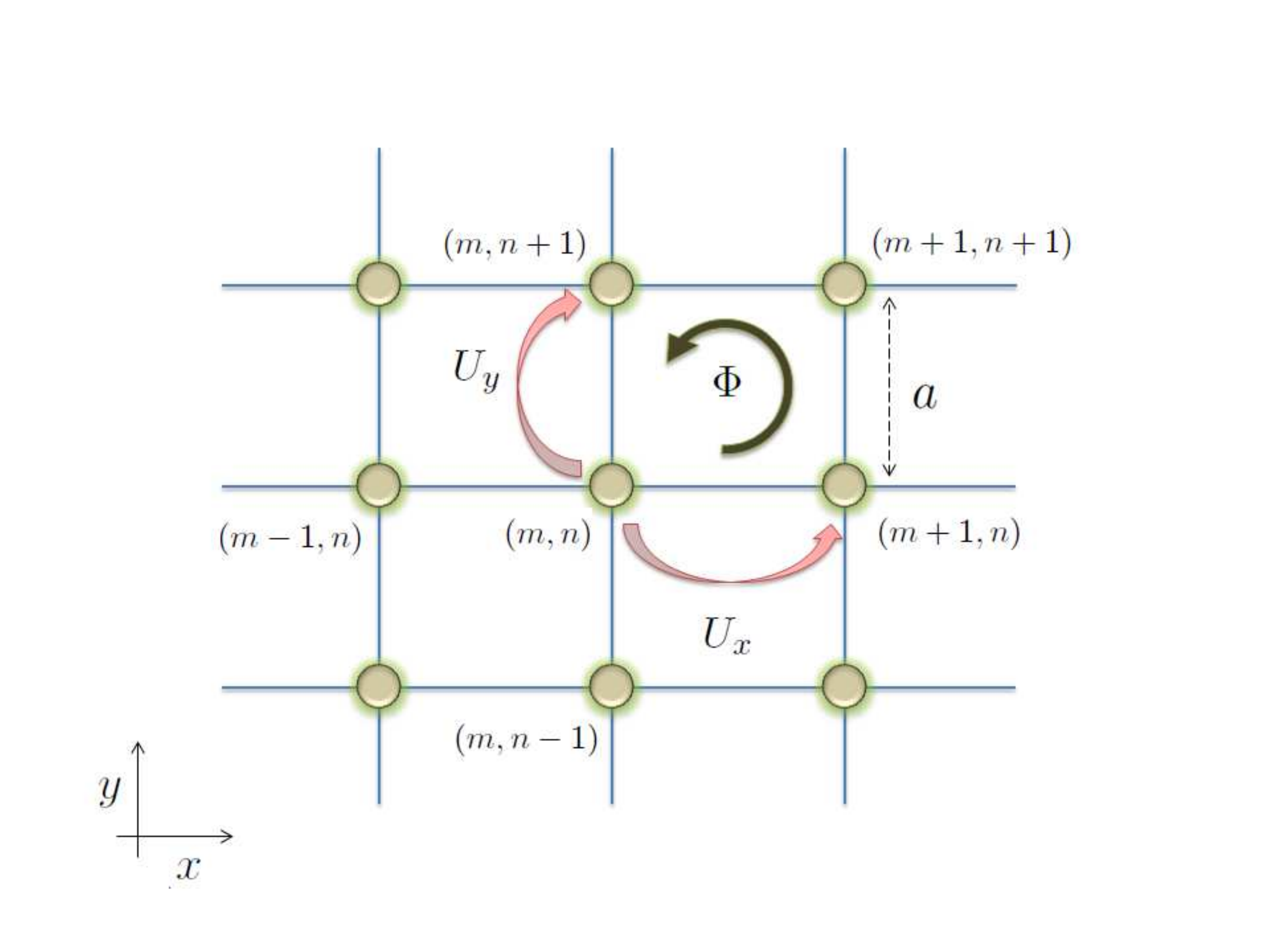} 
\caption{Schematic of an optical lattice. The phase operator $U_x$ determines 
the phase acquired by an atom when it hops from site $(m,n)$ to the site 
$(m+1,n)$. Similarly, the operator $U_y$ determines the phase acquired by 
hopping along the positive y-axis. The operators $U_x^\dag$ and $U_y^\dag$ 
determine the phase acquired in hopping along negative x and y axes, 
respectively. }
\label{Lattice} 
\end{figure}

In this appendix we obtain the full form of the hopping operator $\hB$ in terms 
of the atom creation (annihilation) operators, $\hb^\dag_{(m,n)}$ 
($\hb^{\pdag}_{(m,n)}$). Then we diagonalize it to obtain the spectrum of the tight-binding Hamiltonian in \eqref{eq:TB}. In the end we show how to evaluate the expectation values of this hopping operator with respect to various magnetic orders. 

The lattice sites are indexed as $(m, n)$ and $m,n 
\in \{0,K-1\}$, which makes the lattice a $K \times K$ one (see Figure 
\ref{Lattice}). We also use $i$ and $j$ to shorten the notation for $(m,n)$ and $(m',n')$, respectively. An operator of the form $\hb^\dag_{\sigma, j}\hb^{\pdag}_{\sigma ', i)}$ creates an atom of pseudo-spin $\sigma$ at site $j$ by annihilating an atom of pseudo-spin $\sigma '$ at site $i$. In Figure \ref{Lattice} we have shown the action of all possible hopping operators with non-trivial actions. In presence of a gauge potential, as the particle moves in the lattice potential  its wave function acquires a geometric phase as a result of Aharonov-Bohm  effect. The phase acquired by an atom in hopping from site $\bs r_i$ to $\bs  r_j$, $\phi_{ij}$ is given by
\bea
\phi_{ij} = \int_{\bs r_i}^{\bs r_j} \bs A(\bs r') \cdot d \bs l = \alpha  
\sigma_y (x_j - x_i) \nn \\
 + \beta \sigma_x (y_j - y_i)  + \mathbb{1} B_0 x_i (y_j - y_i).
\eea
For hopping along the x-axis, i.e. $m \rightarrow m\pm1$ the phase acquired is $ \phi_x =  \alpha \sigma_y (x_{i+1} - x_i) + 0 = \alpha \sigma_y $ and for hopping along the y-axis, i.e. $n \rightarrow n\pm1$ it is $\phi_y = 0 + (\beta \sigma_x + B_0 x_i) (y_{i+1} - y_i) = (- \beta \sigma_x + \mathbb{1} B_0 x) $. 

An alternative way to discuss this is to define a set of unitary operators along x and y axes which when act on the wave wave function would produce non-trivial phases. These guage potential dependent phase operators are
\bea
U_x = e^{-i \phi_x}, \quad U_y = e^{-i \phi_y} .
\eea
With our particular choice of vector potential, i.e. $\bs A = (\alpha \sigma_y, 
\beta \sigma_x + 2\pi \Phi m, 0)$ one can calculate the phase operators as:
\bea
U_x = \bem
cos \alpha & - sin \alpha \\
sin \alpha & cos \alpha \\
\eem , 
U_y = e^{-i 2\pi \Phi m} \bem
cos \beta & -i sin \beta \\
i sin \beta & cos \beta \\
\eem .
\eea
Thus a generic form of the tunneling operator $\hB$  (for a 2D lattice) can now 
be written as  
\bea
\hB &= & \sum_{m,n} \hb^\dag_{m+1} U_x \hb_{m} + \hb^\dag_{n+1} U_y \hb_{n} + 
\text{h.c.} .
\eea
Here we have denoted $\hb^\dag_{m}$ for $(\hb^\dag_{m, \upa}$ $\hb^\dag_{m, 
\dna})$, and similarly $\hb^\dag_{n}$ for $(\hb^\dag_{n, \upa}$ $\hb^\dag_{n, 
\dna})$. For our choice of gauge potential we can simplify this equation to
\bea
\hB &=& \sum_{i = x,y} \hB_i^{D} + \hB_i^{ND},  \nn \\
\hB_y^{D} &=& \cos \beta \sum_{n} ( \hb^\dag_{n+1, \upa} \hb^{\pdag}_{n,\upa} + 
\hb^\dag_{n+1, \dna} \hb^{\pdag}_{n,\dna}) e^{- i 2 \pi \Phi m } + \text{h.c.}, 
\nn \\
\hB_y^{ND} &=& -i \sin \beta \sum_{n} ( \hb^\dag_{n+1, \dna} 
\hb^{\pdag}_{n,\upa} + \hb^\dag_{n+1, \upa} \hb^{\pdag}_{n,\dna} ) e^{-i 2 \pi 
\Phi m } + \text{h.c.}, \nn \\
\hB_x^{D} &=& \cos \alpha \sum_{m} \hb^\dag_{m+1, \upa} \hb^{\pdag}_{m,\upa} + 
\hb^\dag_{m+1, \dna} \hb^{\pdag}_{m,\dna} + \text{h.c.}, \nn \\
\hB_x^{ND} &=& \sin \alpha \sum_{m} \hb^\dag_{m+1, \dna} \hb^{\pdag}_{m,\upa} - 
\hb^\dag_{m+1, \upa} \hb^{\pdag}_{m,\dna} + \text{h.c.}.
\label{eq:Hopping-Operator}
\eea
Here the operator is separated into diagonal ($\hB^D_i$) and off-diagonal 
($\hB^{ND}_i$) parts and then each of this part is written for both x and y 
axes, considering only nearest-neighbor interaction. The off-diagonal terms in 
the tunneling operator arise because of the SO coupling. We note that the above 
tunneling matrix can be diagonalized or the SO coupling can be eliminated just by a site dependent rotation. For instance 
the following rotation around x-axis at site $i$ diagonalizes the x axis 
tunneling operator by removing SOC :
\beq
\bem \hb_{i, \upa} \\ \hb_{i, \dna} \eem =
\bem \cos \theta_i & -\sin \theta_i \\
	  \sin \theta_i  & \cos \theta_i 
\eem 
\bem \hb'_{i, \upa} \\ \hb'_{i, \dna} \eem . 
\eeq
Here $\theta_{i+1} - \theta_{i} = \alpha - \pi/2$. So switching on SOC is 
equivalent to rotating the site $i$ about x-axis by an angle $-\theta_i$ and 
along with that the hopping amplitude is also renormalized to $J_1 \cos \alpha$.

\subsection{Diagonalization}
\label{app:diagonalization}
The Hamiltonian in the momentum space can be written as $
\hat{H} = \sum_{\bs k} \hat{\Phi}^\dag_{\bs k} \hat{H}_{\bs k} 
\hat{\Phi}^{\pdag}_{\bs k},$ where $\hat{\Phi}_k = (\hat{b}_{\uparrow \bs k}, 
\hat{b}_{\downarrow \bs k})^T$ is the momentum space representation of the 
two-component spinor and the atomic operators are also written in the momentum 
space representation:
\beq
\hat{b}_s (\bs r) = \frac{1}{\sqrt{N_0}} \sum_{\bs k} e^{i \bs k \cdot \bs r} 
\hat{b}_{s \bs k}, 
\hat{b}^\dag _s (\bs r) = \frac{1}{\sqrt{N_0}} \sum_{\bs k} e^{- i \bs k \cdot 
\bs r} \hat{b}^\dag_{s \bs k} .
\eeq
Writing the atomic operators in the momentum basis we can diagonalize the 
Hamiltonian (with out the interaction part) obtained in \eqref{eq:TB},
\bea
\hat{H} = - \mj \sum_{s} \sum_{<m,n>} \frac{1}{N} \sum_{\bs k,\bs k'} 
\hat{b}^\dag_{s \bs k} ( e^{i k_x} e^{- i \sigma_y \alpha} \nn \\ + e^{i k_y} 
e^{-i 2\pi \Phi m} e^{i \sigma_x \alpha} ) \hat{b}_{s \bs k'}  + \text{h.c.}
\eea
Now we invoke orthonormality of plane wave basis: $ \frac{1}{N} \sum_{\bs r} 
e^{- i \bs r \cdot (\bs k - \bs k')} = \delta(\bs k - \bs k'),$ and the Euler's 
identity, $ \exp[i \theta (\hat{n} \cdot \vec{\sigma})] = \bs 1 \cos \theta + i 
(\hat{n} \cdot \vec{\sigma}) \sin \theta $ and denoting $\epsilon_m = k_y - 2 
\pi m \Phi$ we obtain
\bea
\hat{H}_k = \cos\alpha ( \cos \epsilon_m + \cos k_x) \bs 1 \nn \\
- \sin \alpha (\sin\epsilon_m \sigma_x - \imath \sin k_x \sigma_y ). 
\eea
Using the $2\times2$ representation of the Pauli matrices we obtain a $2 \times 2$ Hamiltonian. Writing this Hamiltonian in its eigen-basis we diagonalize it. Thus the spectrum is
\bea
E_\pm = 2 \cos \alpha (\cos \epsilon_m + \cos k_x )  \nn \\
\pm 2 \sin \alpha \sqrt{\sin^2\epsilon_m + \sin^2k_x } 
\eea

\subsection{Expectation Values}
\label{app:tunnel}
In this section we calculate $\bra{\Psi_{MI}}\hB \ket{\Psi_{MI}}$, which appears in equation \eqref{eq:nph}. The full form of $\hB$ is obtained in \eqref{eq:Hopping-Operator}. We assume there are exactly equal number of lattice 
in the A and B sub-lattices, hence the total number of lattice sites is even, 
i.e. $N_0 = K^2$ is even. We demonstrate the calculation for a simple $2\times2$ sites problem and then generalize it for multiple sites. In this case the MI wave function becomes 
\bea
\ket{\Psi_{MI}} = \ket{\psi_A}_{00} \ket{\psi_B}_{01} \ket{\psi_A}_{11} \ket{\psi_B}_{10} .
\eea
The bottom left site is used as the origin of the coordinate system and $(m,n) = (0,0)$ is shortened to $00$, similarly other sites are indexed. Here $\ket{\psi_{A,B}} = \cos \frac{\theta_{A,B}}{2} \ket{\upa} + e^{i \phi_{A,B}} \sin \frac{\theta_{A,B}}{2} \ket{\dna} $. When the operator $\hb^\dag_{m+1, \upa} \hb^{\pdag}_{m,\upa}$ (fixing $n=1$) acts on the above wave then (say, $m=1$) it hops a $\upa$ spin from site $m$ (=1) to $m+1$ (=0). Thus the resulting wave function becomes : 
\bea
\hb^\dag_{m+1, \upa} \hb^{\pdag}_{m,\upa} \ket{\Psi_{MI}} = \ket{\psi_A}_{00} \Big ( \cos \frac{\theta_B}{2} \ket{\uparrow, \uparrow} +  e^{i \phi_B} \sin \frac{\theta_B}{2} \ket{\dna} \Big )_{01} \nn \\ 
\Big ( \cos \frac{\theta_A}{2} \ket{0} +  e^{i \phi_{A}} \sin \frac{\theta_A}{2} \ket{\dna} \Big )_{11} \ket{\psi_B}_{10} .
\eea
Here $\ket{0}$ denotes the spin-vacuum. When $\bra{\Psi_{MI}}$ is acted on the left side of the above expression we obtain
\bea
\bra{\Psi_{MI}}\hb^\dag_{m+1, \upa} \hb^{\pdag}_{m,\upa} \ket{\Psi_{MI}} &=&
\langle \psi_A \ket{\psi_A} \Big ( 0 + \sin^2 \frac{\theta_B}{2} \Big ) \Big ( 0 + \sin^2 \frac{\theta_A}{2} \Big ) \bra{\psi_B} \psi_B \rangle \nn \\
&=& \sin^2 \frac{\theta_A}{2} \sin^2 \frac{\theta_B}{2} .
\eea
The hermitian conjugate of this operator hops $\upa$ from $m+1$ to $m$. Thus 
$\bra{\Psi_{MI}} (\hb^\dag_{m+1, \upa} \hb^{\pdag}_{m,\upa} + h.c. ) \ket{\Psi_{MI}} = 2\sin^2 \frac{\theta_B}{2} \sin^2 \frac{\theta_A}{2}.$ In a similar way we can obtain
\bea
\bra{\Psi_{MI}}\hb^\dag_{m+1, \dna} \hb^{\pdag}_{m,\dna} \ket{\Psi_{MI}} = \cos^2 \frac{\theta_A}{2} \cos^2 \frac{\theta_B}{2} .
\eea
Now for hoppings associated with spin flip can be obtained as:
\bea
\hb^\dag_{m+1, \dna} \hb^{\pdag}_{m,\upa} \ket{\Psi_{MI}} 
&=& \ket{\psi_A}_{00} \Big ( \cos \frac{\theta_B}{2} \ket{\uparrow} +  e^{i \phi_B} \sin \frac{\theta_B}{2} \ket{\dna, \upa} \Big )_{01} \nn \\ 
&& \Big ( \cos \frac{\theta_A}{2} \ket{0} +  e^{i \phi_{A}} \sin \frac{\theta_A}{2} \ket{\dna} \Big )_{11} \ket{\psi_B}_{10} \nn \\
&=& 0.
\eea

So terms like  $\hb^\dag_{m+1, \dna} \hb^{\pdag}_{m,\upa}$ or, $\hb^\dag_{m+1, \upa} \hb^{\pdag}_{m,\dna}$ don't contribute to the expectation. When we have a $K \times K$ lattice there will be $K-1$ hopping possible along x-axis yielding a contribution of $2(K-1) \cos^2 \frac{\theta_A}{2} \cos^2 \frac{\theta_B}{2} + \sin^2 \frac{\theta_B}{2} \sin^2 \frac{\theta_A}{2}$. There are $K$ such x-axes so total contribution becomes 
\bea
\expect{\hB_{x}} = 2 \cos \alpha K(K-1) \Big [ \sin^2\frac{\theta_A}{2} 
\sin^2\frac{\theta_B}{2} \nn \\ 
 + \cos^2\frac{\theta_A}{2} \cos^2\frac{\theta_B}{2} \Big]
\eea
Now we turn to hopping along y-axis. We switch on the Abelian gauge field discussed in the main-text, see equation \eqref{eq:Hopping-Operator} for the full form of the Hopping operator. Hence now each hopping along y-axis is associated with a phase depending upon the x-axis coordinate of the site, i.e. $e^{-2\pi i \Phi m}$. For hopping along -y the phase is $e^{+ 2\pi i \Phi m}$. Using a similar argument we arrive at the following result :
\bea
\expect{\hB_{y}} = 2 \cos \beta (K-1) \sum_{m = 0}^{K-1} \cos(2\pi m \Phi) \nn 
\\ \Big [\sin^2\frac{\theta_A}{2} \sin^2\frac{\theta_B}{2} + 
\cos^2\frac{\theta_A}{2} \cos^2\frac{\theta_B}{2}] .
\eea
The last expression can be simplified to $f(K, \Phi) = \sum_{m = 0}^{K-1} 
\cos(2\pi m \Phi) = \frac{\sin(K \pi \Phi)}{\sin(\pi \Phi)} \cos[\pi \Phi 
(K-1)]$. Thus the full expectation becomes,
\bea
\expect{\hB} = 2 \cos \alpha (K-1)\Big [\cos \alpha K + \cos \beta f(K, \Phi) 
\Big] \nn \\ \Big[\sin^2\frac{\theta_A}{2} \sin^2\frac{\theta_B}{2} + 
\cos^2\frac{\theta_A}{2} \cos^2\frac{\theta_B}{2} \Big].
\eea

\end{document}